\documentclass[12pt]{article}
\usepackage{amsmath,amssymb,amsthm,amscd}
\usepackage{tikz}
\usepackage{jheppub}

\newcommand{\CF}{{\cal F}}
\newcommand{\cF}{\mathcal{F}}

\newcommand{\cM}{{\mathcal{M}}}
\newcommand{\CN}{{\cal N}}
\newcommand{\cN}{{\mathcal{N}}}

\newcommand{\CT}{{\cal T}}
\newcommand{\cT}{\mathcal{T}}

\newcommand{\CW}{{\cal W}}

\newcommand{\IC}{\mathbb{C}}
\newcommand{\IP}{\mathbb{P}}

\newcommand{\bR}{\mathbb{R}}

\newcommand{\IZ}{\mathbb{Z}}

\newcommand{\fm}{\mathfrak{m}}
\newcommand{\M}{\mathfrak{M}}
\newcommand{\fM}{\mathfrak{M}}
\newcommand{\fN}{\mathfrak{N}}

\newcommand{\tm}{\tilde{m}}
\newcommand{\qt}{\tilde{q}}
\newcommand{\Qt}{\widetilde{Q}}

\newcommand{\wt}{\widetilde}
\newcommand{\wh}{\widehat}

\DeclareMathOperator{\rank}{rank}
\DeclareMathOperator{\Tr}{Tr}

\newcommand{\rep}[1]{\ensuremath{\mathbf{#1}}}
\newcommand{\ud}[2]{^{#1}_{\phantom{#1}#2}}

\newcommand{\nn}{\nonumber}
\newcommand{\eg}{\textit{e.g.}}
\newcommand{\ie}{\textit{i.e.}}
\newcommand{\be}{\begin{equation}}
\newcommand{\ee}{\end{equation}}
\newcommand{\ben}{\begin{eqnarray}\displaystyle}
\newcommand{\een}{\end{eqnarray}}
\newcommand{\bea}{\begin{equation}\begin{aligned}}
\newcommand{\eea}{\end{aligned}\end{equation}}
\newcommand{\smat}[1]{{\left( \begin{smallmatrix} #1 \end{smallmatrix} \right)}}

\newdimen\tableauside\tableauside=1.0ex
\newdimen\tableaurule\tableaurule=0.4pt
\newdimen\tableaustep
\def\phantomhrule#1{\hbox{\vbox to0pt{\hrule height\tableaurule width#1\vss}}}
\def\phantomvrule#1{\vbox{\hbox to0pt{\vrule width\tableaurule height#1\hss}}}
\def\sqr{\vbox{%
  \phantomhrule\tableaustep
  \hbox{\phantomvrule\tableaustep\kern\tableaustep\phantomvrule\tableaustep}%
  \hbox{\vbox{\phantomhrule\tableauside}\kern-\tableaurule}}}
\def\squares#1{\hbox{\count0=#1\noindent\loop\sqr
  \advance\count0 by-1 \ifnum\count0>0\repeat}}
\def\tableau#1{\vcenter{\offinterlineskip
  \tableaustep=\tableauside\advance\tableaustep by-\tableaurule
  \kern\normallineskip\hbox
    {\kern\normallineskip\vbox
      {\gettableau#1 0 }%
     \kern\normallineskip\kern\tableaurule}%
  \kern\normallineskip\kern\tableaurule}}
\def\gettableau#1{\ifnum#1=0\let\next=\null\else
\squares{#1}\let\next=\gettableau\fi\next}

\preprint{SISSA  11/2017/MATE-FISI, 19/2017/FISI}

\title{SUSY monopole potentials in 2+1 dimensions}

\author[1,2]{Francesco Benini}
\author[2]{Sergio Benvenuti}
\author[3]{Sara Pasquetti}

\affiliation[1]{Institute for Advanced Study, Princeton, NJ 08540, USA}
\affiliation[2]{International School of Advanced Studies (SISSA) \& INFN, Sezione di Trieste \\
via Bonomea 265, 34136 Trieste, Italy}
\affiliation[3]{Dipartimento di Fisica, Universit\`a di Milano-Bicocca \& INFN, Sezione di Milano-Bicocca, \\
I-20126 Milano, Italy
}
\emailAdd{fbenini@sissa.it,benve79@gmail.com,sara.pasquetti@gmail.com.}

\abstract{Gauge theories in 2+1 dimensions can admit monopole operators in the potential. Starting with the theory without monopole potential, if the monopole potential is relevant there is an RG flow to the monopole-deformed theory. Here, focusing on $U(N_c)$ SQCD with $N_f$ flavors and $\CN=2$ supersymmetry, we show that even when the monopole potential is irrelevant, the monopole-modified theory $\mathcal{T}_\mathfrak{M}$ can exist and enjoy Seiberg-like dualities. We provide a renormalizable UV completion of $\mathcal{T}_\mathfrak{M}$ and an electric-magnetic dual description $\mathcal{T}_\mathfrak{M}'$. We subject our proposal to various consistency checks such as mass deformations and $S^3_b$ partition functions checks. We observe that $\mathcal{T}_\mathfrak{M}$ is the S-duality wall of 4D $\cN=2$ SQCD. We also consider monopole-deformed theories with Chern-Simons couplings and their duals.
}

\begin{document}
\maketitle

\section{Introduction and results}

Three-dimensional gauge theories admit an interesting class of gauge-invariant disorder operators which can be defined by prescribing suitable boundary conditions around a point for the gauge fields in the path integral. These operators carry a magnetic (or topological) charge, hence they are called monopole operators and create some units of  magnetic flux on a two-sphere surrounding their insertion point. Despite being local, these  operators are not polynomial in the elementary fields and this fact makes it difficult  to study what happens when they are added to the Lagrangian.

Long ago Polyakov \cite{Polyakov:1976fu} showed that monopole operators can actually appear in the potential, at the infrared fixed-point of an RG-flow triggered by Higgsing a gauge symmetry leading to confinement.
 
In condensed-matter physics, quantum mechanical lattice models in two spatial dimensions admit, in the thermodynamic limit, interesting second-order quantum phase transitions which should be described by a $(2+1)$-dimensional conformal field theory (CFT) \cite{Senthil:2004aza}. Such a CFT$_3$ could for instance be a $U(1)$ gauge theory with some number of fermionic and/or scalar charged matter fields (flavors). A natural question is if the potential of the gauge theory description contains monopole operators $\M$ or not. CFTs with monopoles admit a smaller global symmetry, but on the lattice it is not easy to understand what the emergent low-energy global symmetries are, and so it is not clear whether the monopole potential is generated or not. Much work has been devoted to investigate this question, see for instance \cite{Dyer:2015zha, lou2009antiferromagnetic, kaul2012lattice, block2013fate} for some examples.

Some of that work has focused on trying to determine the scaling dimension $\Delta[\M]$ of monopole operators $\M$ in the infra-red (IR) of the gauge theory without monopole potential. We will call such a IR theory $\CT_0$. If the scaling dimension $\Delta[\M]$ is below $3$ in $\CT_0$, then the monopole deformation is relevant and we naturally expect the monopole potential to be turned on in the absence of fine-tuning. This triggers an RG flow to some other phase. Usually the theory has some other relevant parameter that, as varied, leads the RG flow to different phases. If the phase transition is second order, one can tune the relevant parameter and obtain a fixed point $\CT_\M$, different from $\CT_0$. Of course, it might well happen that the phase transition is first order and the fixed point $\CT_\M$ does not exist. In general answering these questions is very hard.

In this paper we study the effect of adding monopole operators to the Lagrangian of supersymmetric (SUSY) theories. This simplifies our lives because supersymmetry gives us much more non-perturbative control on the dynamics of the theories. In various cases we can argue that the phase transitions are second order, and so we can argue for the existence of the CFTs $\CT_\M$.

Monopole operators in SUSY theories have been  extensively investigated (see \eg{} \cite{borokhov2003topological, borokhov2003monopole, Gaiotto:2008ak, Gaiotto:2009tk, Kim:2009wb, Benna:2009xd, Benini:2009qs, Jafferis:2009th, Bashkirov:2010kz, Imamura:2011su, Benini:2011cma, Benini:2011mf, Cremonesi:2013lqa, Cremonesi:2014kwa, Cremonesi:2014vla}). Supersymmetry allows for a quantitative control over the scaling dimensions of supersymmetric operators and in particular of supersymmetric monopoles. This in turn makes it possible to study in great details the moduli space of vacua and to determine how the gauge invariant operators map when multiple dual descriptions of the same IR physics---so-called IR dualities---are available.

Superpotentials involving monopole operators have appeared in the literature in various circumstances. For example, they famously appear in the Aharony dual of  $U(N_c)$ or $USp(2N_c)$ theories with fundamental quarks \cite{Aharony:1997gp}. In  \cite{Intriligator:2013lca} it has been shown how to obtain the Aharony pair of  $U(N_c)$ dual theories starting from a duality with Chern-Simons couplings and no monopole superpotentials, by turning on suitable real mass deformations.  In this process the monopoles enter the superpotential  as 3D  instanton effects \cite{Affleck:1982as}, so this construction provides a UV completion of the Aharony dual pairs in terms of renormalizable theories.
 
Monopole superpotentials can also appear as the effect of reducing a 4D theory on a circle down to 3D \cite{Aharony:1997bx, Aharony:1997gp, Aharony:2013dha, Aharony:2013kma}. A careful study of the moduli spaces indicates that, contrary to the naive dimensional  reduction, the compactification on a circle of finite size allows for the generation of Kaluza-Klein monopoles which enter the superpotential. These monopoles play a key role in consistently deriving 3D dualities from 4D ones.  When reducing on a circle a pair of dual 4D theories, at the first step one obtains a 3D dual pair with monopole superpotentials. The monopole operators are charged under topological and axial symmetries and  break these symmetries in 3D (such symmetries would be anomalous or non-existent in 4D). At this point one can turn on various real mass deformations and recover 3D dualities without monopole superpotentials. This procedure has been successfully  implemented for theories with various gauge and matter content, for a review see \cite{Amariti:2016kat} and references therein.

Rather than trying to get rid of the monopoles, one can also turn on other real mass deformations, flow to new theories with monopole superpotentials and perhaps discover new dualities. This is what we do in this paper. Starting from the  4D Intriligator-Pouliot duality \cite{Seiberg:1994pq, Intriligator:1995ne} we arrive at $U(N_c)$ SQCD gauge theories with $N_f$  flavors and superpotential $\CW_\text{mon}=\M^+ + \M^-$, where  $\M^\pm$ are the two simplest supersymmetric monopoles in the theory, with topological charge $\pm 1$.  We call this theory $\CT_\M$ and propose that it has a dual description $\CT'_\M$ given by a $U(N_f - N_c - 2)$ theory with $N_f$ flavors, $N_f^2$ gauge singlets and superpotential $\CW = \sum_{ij=1}^{N_f} M\ud{i}{j} \tilde q_i q^j+ \widehat\M^+ + \widehat\M^-$.  We provide various evidences of this duality, matching the gauge-invariant operators, comparing partition functions and performing real and complex mass deformations.

There are various reasons why we are interested in theories with monopole superpotentials turned on. They appear in the context of  the 3D-3D correspondence of \cite{Dimofte:2011ju, Dimofte:2011py, Kashaev:2012cz, Teschner:2012em, Dimofte:2012pd, Dimofte:2013lba}, in brane setups with low enough supersymmetry \cite{Benvenuti:2016wet} and in T-brane systems \cite{Collinucci:2016hpz}. It also seems that monopole superpotentials appear in theories describing certain 3D duality walls  \cite{Floch:2015hwo} and codimension two defects in 5D \cite{Gomis:2014eya, Pan:2016fbl}.
The sphere partition functions of those domain walls and defect theories can be obtained using the AGT correspondence \cite{Alday:2009aq}. The match to the CFT calculations often requires tuning the real mass and FI parameters of the gauge theory to some specific values. The presence of monopoles operators in the superpotential can explain these tunings. For example, we show that this is the case for the SQCD $S$-duality wall which we identify with our theory $\CT_\M$, for $N_f = 2 (N_c+1)$.

\begin{figure}[t]
\begin{center}
\includegraphics[width=3in]{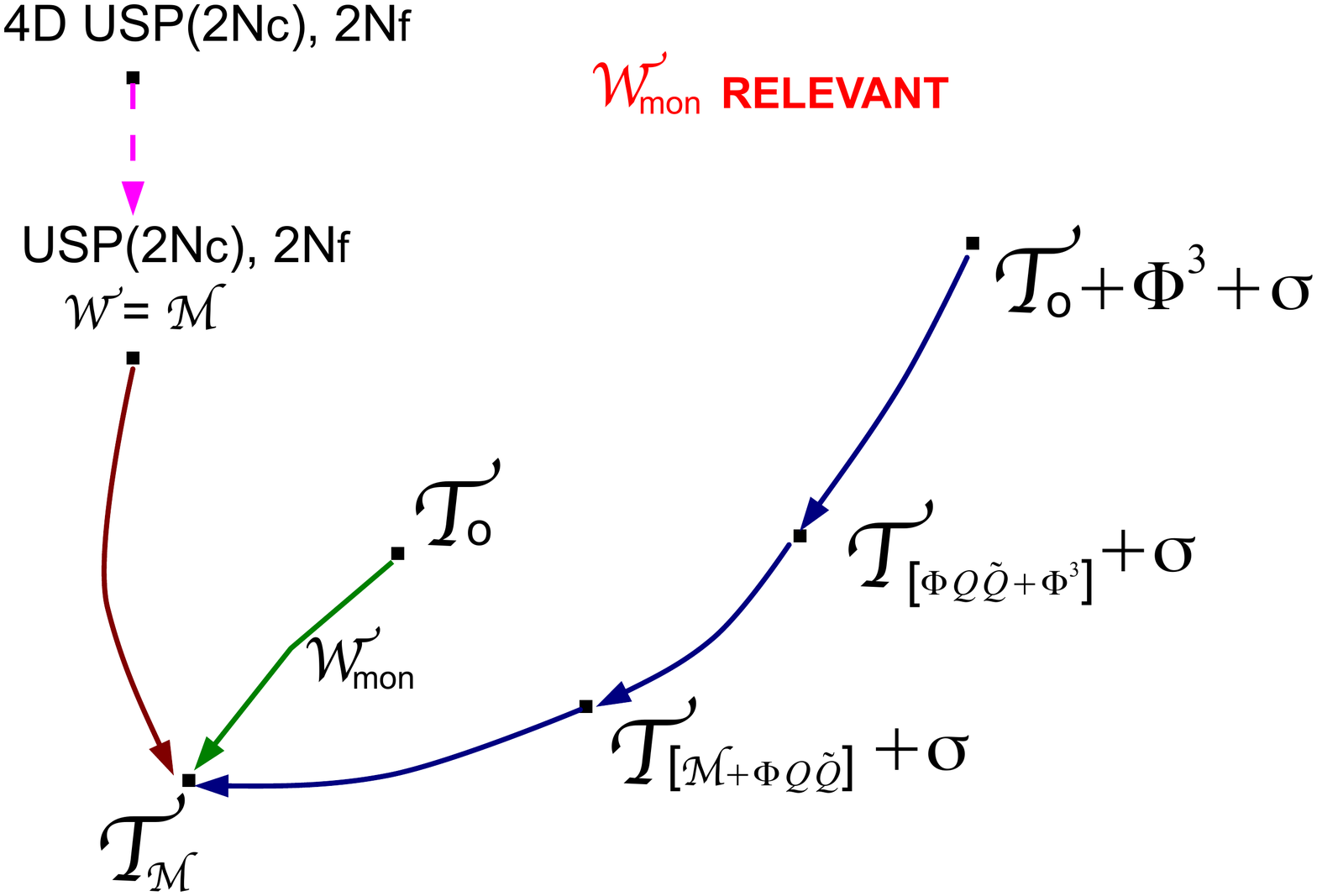} \hspace{\stretch{1}} \includegraphics[width=3in]{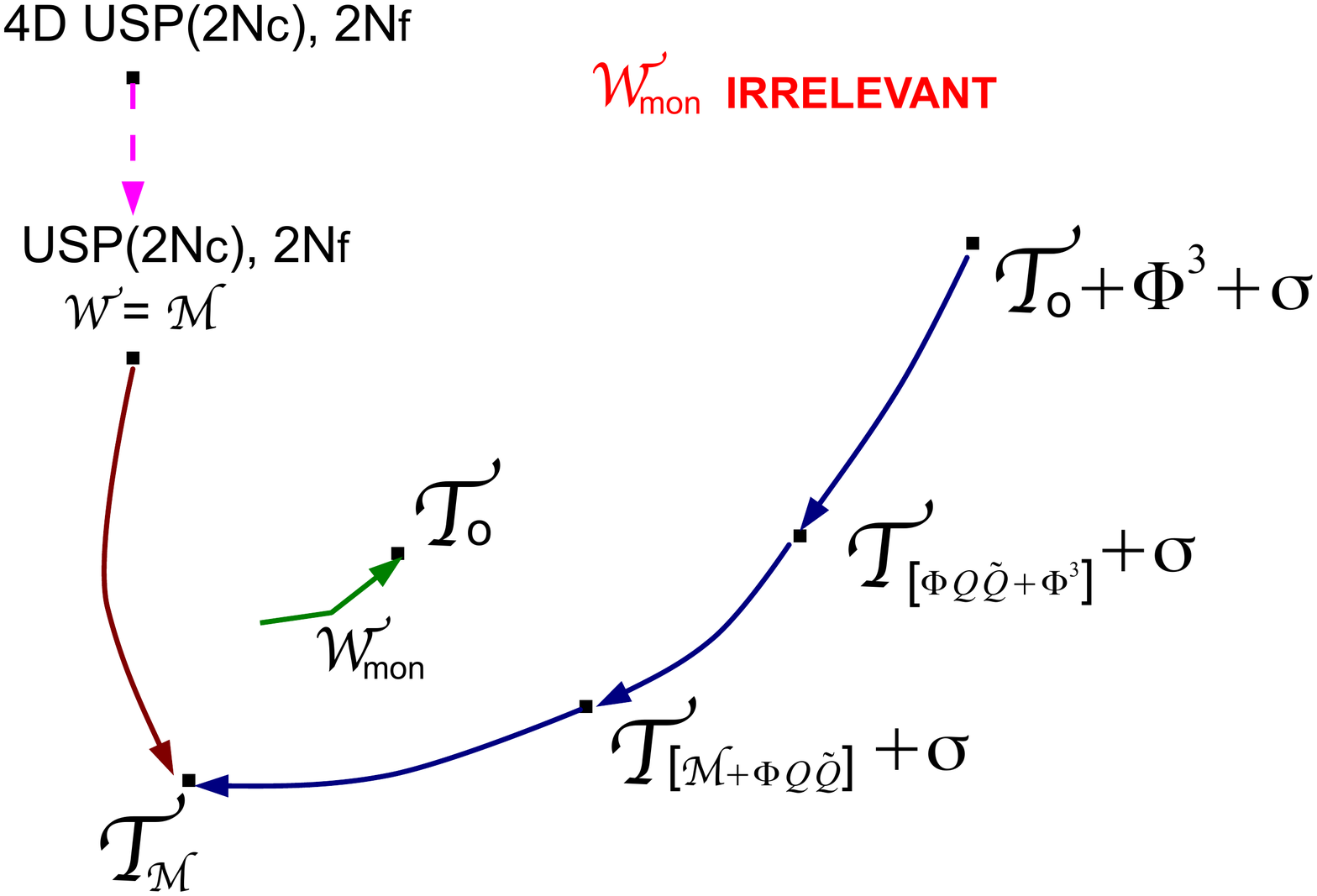}
\caption{Schematic structure of RG flows that can lead to $\CT_\fM$. The pink flow is a $\text{4D}\to\text{3D}$ reduction on $S^1$, while the red flow is a real mass deformation. The green flow is a deformation of $\CT_0$ by $\CW_\text{mon}$, which is relevant on the left but irrelevant on the right (and therefore it does not leave $\CT_0$). The blue flow involves more degrees of freedom: the Ising-SCFT and extra free fields (see Section \ref{superisi}).
\label{FIGintro}}
\end{center}
\end{figure}

One of the most crucial questions is whether the fixed point $\CT_\M$ exists---as we vary $N_c, N_f$---as a CFT distinct from $\CT_0$ (the latter is the fixed point of the theory with no monopole superpotential). The $\text{4D} \to \text{3D}$ construction involving the circle reduction of the 4D Intriligator-Pouliot duality followed by a real mass deformation (schematically depicted as the pink and red flows in Fig.~\ref{FIGintro}) does not really answer this question. For instance, in a range of values of $N_c, N_f$ it might well happen that $\CT_\M$ is an irrelevant deformation of $\CT_0$ (the monopole superpotential is a dangerously irrelevant deformation). If the fixed point $\CT_\M$ exists, another question is what RG flows can reach it.

The conservative approach to answer those questions, as mentioned earlier, is to first flow to the fixed point $\CT_0$ of the $U(N_c)$ SQCD without the monopole superpotential, and then try to reach $\CT_\M$ (this corresponds to the green flow in Fig.~\ref{FIGintro} on the left). This flow is possible  when $\CW_\text{mon}$ is a relevant deformation in  $\CT_0$ which, for any fixed value of $N_c$, only happens for two or three values of $N_f$ \cite{Safdi:2012re}. If $N_f$ is larger, $\CW_\text{mon}$ is a dangerously irrelevant operator which deforms the moduli space but cannot trigger  the flow to $\CT_\M$ (Fig.~\ref{FIGintro} on the right).

However one could  try to reach $\CT_\M$ with other more involved  flows. For example, we will show that starting from $\CT_0$ plus some decoupled copies of the so-called Ising-SCFT and other free fields, turning on suitable couplings we can reach $\CT_\M$ for $N_f$ up to $3N_c+3$. This chain of flows is represented by the blue arrows in Fig.~\ref{FIGintro}. In other words, we can argue for the existence of $\CT_\M$ in a much wider window than the narrow one of relevance of $\CW_\text{mon}$ in $\CT_0$.  

The results of this paper strongly suggest that, even in cases when $\CW_\text{mon}$ is irrelevant in $\CT_0$, $\CT_\M$ might be the thermodynamic limit of quantum spin models on a planar lattice. This might  be relevant also for non-supersymmetric models.

As we explore the parameter space $(N_c, N_f)$ and lower the number of flavors below $N_f=3N_c+3$ we find a very rich structure. For example, when we  cross the  unitarity bound for mesonic operators we encounter some decoupled sectors. Continuing to lower the number of flavors at fixed $N_c$ we encounter, at $N_f = N_c+2$, an effective description in terms of a Wess-Zumino model, at $N_f = N_c+1$ a smooth quantum-deformed moduli space, and for $N_f\leq N_c$ no vacua.

We do not know if the fixed point $\CT_\M$ still exists for values of $N_f$ larger than $3N_c+3$, and in that case whether there is an upper bound on the value of $N_f$ and what this upper bound could be. We leave this important issue for future work.

Starting from the $\CT_\M = \CT'_\M$ duality and turning on suitable real mass deformations, we obtain various other dualities. For instance, dualities for theories with a single monopole in the superpotential, say $\CW = \M^+$, and possibly with Chern-Simons couplings. We also consider higher monopole deformations, $\CW = (\M^+)^n + (\M^-)^n$ with $n=2,3$.

It is clear that the analysis in this paper could be repeated for theories with more general gauge group and matter content. It would of course be very interesting to know quantitatively what happens in non-supersymmetric examples.

\

The paper is organized as follows. 
In Section \ref{sec: IRphysics} we introduce the theory $\CT_\M$ and its dual $\CT'_\M$ by reducing the 4D Intriligator-Pouliot duality \cite{Seiberg:1994pq, Intriligator:1995ne} on a circle. We describe some basic properties of the SCFT $\CT_\M$, discuss a unitarity bound on $N_f$ and match the moduli space of vacua to that of $\CT'_\M$. In Section \ref{Dynamics} we describe the dynamics of $\CT_\M$ as we vary $N_c$ and $ N_f$.

In Section \ref{sec: RGflows} we describe how to obtain $\CT_\M$ from an RG flow that starts from a 3D weakly-coupled renormalizable theory, when $N_f \leq 3N_c + 3$. This involves adding more degrees of freedom.

In Section \ref{sec:dualitywall} we relate $\CT_\M$ with $N_f = 2N_c+2$ to the duality-wall theory of 4D $\CN=2$ SQCD found in \cite{Floch:2015hwo}.

In Section \ref{sec: higher powers} we briefly explore the case of higher-monopole superpotentials.

In Section \ref{partfun} we derive, at the level of the $S^3_b$ partition function, the $\text{4D} \to \text{3D}$ flow that leads to the duality $\CT_\M=\CT'_\M$. We also show that our duality reduces to the Aharony duality upon a suitable real mass deformation.

In Section \ref{newonemon} we consider more real mass deformations of the duality $\CT_\M=\CT'_\M$. We derive new dualities involving a superpotenital $\CW=\M^+$, and dualities for theories with Chern-Simons couplings.


\section{$\boldsymbol{U(N_c)}$ SQCD with monopole superpotential and its dual}
\label{sec: IRphysics}

In this section we introduce our main characters: the theory $U(N_c)$ SQCD with linear monopole superpotential $\CW_\text{\rm mon}$, and the fixed point $\CT_{\M}$.  Monopole operators are local disorder operators which create magnetic flux on the two-sphere surrounding the insertion point.
In recent years there has been much progress in understanding the properties of these operators in 3D $\CN=4$ and $\CN=2$ gauge theories, see \eg{} \cite{borokhov2003topological, borokhov2003monopole, Gaiotto:2008ak, Gaiotto:2009tk, Kim:2009wb, Benna:2009xd, Benini:2009qs, Jafferis:2009th, Bashkirov:2010kz, Imamura:2011su, Benini:2011cma, Benini:2011mf, Cremonesi:2013lqa, Cremonesi:2014kwa, Cremonesi:2014vla}. For example it has been derived a formula to compute the charge 
of a monopole operator of magnetic charge $\fm$ under any Abelian global symmetry. 
The quantum corrections $\delta q$ to the charges of monopoles are obtained via the one-loop formula
\be
\delta q[\M] = - \frac12 \sum_{\text{fermions }\psi} q(\psi) \, \big| \rho_\psi(\fm) \big| \;,
\label{monfor}
\ee
where the fermions $\psi$ transform as the weights $\rho_\psi$ under the gauge group.

In the case of SQCD with $N_f$ flavors and vanishing superpotential $\CW=0$ there are two fundamental monopole operators $\M^+$, $\M^-$ that correspond to magnetic fluxes $\fm = (1, 0, \dots, 0)$ and $\fm = (0, \dots, 0, -1)$. The continuous global symmetry of the theory is $U(1)_R \times U(1)_A \times SU(N_f)^2 \times U(1)_T$ and the table of charges is the following:
\be
\begin{array}{c|cccccc}
 & U(1)_R & \quad & U(1)_A\,\, & \, SU(N_f)_\ell \, & \, SU(N_f)_r \, & \, U(1)_T \\
\hline
Q & R_Q & & 1 & \rep{N_f} & \rep{1} & 0 \\
\Qt & R_Q & & 1 & \rep{1} & \overline{\rep{N_f}} & 0 \\
\hline
M & 2 R_Q & & 2 & \rep{N_f} & \overline{\rep{N_f}} & 0 \\
\M^+ & (1-R_Q)N_f - N_c + 1 & & - N_f & \rep{1} & \rep{1} & 1 \\
\M^- & (1-R_Q)N_f - N_c + 1 & & - N_f & \rep{1} & \rep{1} & -1
\end{array}
\ee
The monopoles $\M^\pm$ have charges $(\pm1 , - N_f)$ under the topological and axial symmetries $U(1)_T \times U(1)_A$.
We will first consider the superpotential
\be
\label{Wmon}
\CW_\text{\rm mon} = \M^+ + \M^-\,.
\ee
Notice that (\ref{Wmon}) breaks both $U(1)_T$ and $U(1)_A$, but it does not break the discrete $\IZ_2$ charge conjugation symmetry. In Section \ref{sec: higher powers} we will study other charge conjugation symmetric choices, while in Section \ref{newonemon} we will consider adding only $\M^+$ or $\M^-$, which preserves one combination of $U(1)_T$ and $U(1)_A$ but breaks charge conjugation. 
The monopole superpotential $\M^+ \M^-$ in the $U(N_c)$  theory has been discussed in \cite{Aharony:2013dha}.

\subsection[$\cT_\fM$ and its dual from 4D]{$\boldsymbol{\cT_\fM}$ and its dual from 4D}\label{4d3dflow}

A possible way to reach $\cT_\fM$ is to start from the 4D $USp(2N_c)$ SQCD theory with $2N_f$ fundamental flavors. This theory has a dual description as a $USp(2N_f - 2N_c - 4)$ theory with $2N_f$ fundamental flavors, $N_f(2N_f-1)$ singlets $M_{ab}$ organized into an antisymmetric matrix, and superpotential $\CW = \sum_{a<b}^{2N_f} M^{ab} q_a {\cdot} q_b$ \cite{Seiberg:1994pq, Intriligator:1995ne}. When this dual pair is compactified on $\bR^3 \times S^1$ \cite{Aharony:1997bx, Aharony:2013dha}, non-perturbative effects due to Euclidean monopole configurations wrapping the circle generate superpotential terms proportional to the monopole operators. On the electric side one finds that $\CW = \eta \, \M$ is generated, while on the magnetic side one finds $\tilde\eta \, \widehat \M$, where $\M$, $\widehat \M$ are the monopole operators parameterizing the Coulomb branches of the electric and magnetic theory, respectively, and $\eta$, $\tilde\eta$ are energy scales. In particular $\eta = \Lambda^b$ with $b$ the one-loop beta function and $\Lambda = \mu\exp \big( -\tfrac{4\pi}{g_4(\mu)^2} \big)$ the dynamically generated scale. In the rest of this paper we will mostly omit the coefficients $\eta$, $\tilde{\eta}$. The deformation by the monopole superpotential drives the theory to a non-trivial fixed point.
 
This duality was tested in \cite{Aharony:2013dha} at the level of the partition function on the squashed three-sphere $S^3_b$ \cite{Hama:2011ea}. Since the monopole superpotential breaks the $U(1)_A$ symmetry, we cannot turn on the real mass deformation associated to it. At the level of the $S^3_b$ partition function this fact appears as a constraint on the mass parameters:
\be
\label{costraing}
\sum\nolimits_{a=1}^{2N_f} m_a= iQ (N_f-N_c-1) \;,
\ee
where $Q=b^2+b^{-2}$ and $b$ is the squashing parameter. 

In \cite{Aharony:2013dha}  it was also  shown how to recover the 3D Aharony duality for $USp(2N_c)$ theories \cite{Aharony:1997gp}  from this compactified 4D duality. The idea is to start with $2N_f+2$ flavors and take the real mass deformation $m_{2N_f+1}=s+\alpha$, $m_{2N_f+2}=-s+\alpha$ with $s\to\infty$. Since  $\alpha$ is a free parameter  the constraint (\ref{costraing}) is lifted and $U(1)_A$ is restored. The only SUSY vacuum on both sides of the duality is the trivial one, in which the real scalar in the vector multiplet (which can be diagonalised by a gauge rotation) takes zero VEV, $\langle\sigma_j \rangle=0$.
On the electric side the limit reduces the theory to $USp(2N_c)$ with $2N_f$ flavors and no superpotential. On the magnetic side the limit gives $USp(2N_f-2N_c-2)$ with $2N_f$ flavors. The limit reduces the meson matrix to a $N_f(2N_f-1)$ block $M^{ab}$ plus an extra singlet $S$ which couples linearly to the dual monopole $\widehat \M$ in the magnetic superpotential $\CW = \sum_{a<b}^{2N_f} M^{ab} q_a {\cdot} q_b+ S \, \widehat \M$.

Here we are interested in  a different 3D limit. We split the $2N_f$ masses into two sets, $m_1, \dots, m_{N_f}$ and $\tilde m_1, \dots, \tilde m_{N_f}$ and consider the real  mass deformation
\be
\label{rmd}
m_a \to m_a +s \;,\qquad \tilde m_b \to \tilde m_b -s \;, \qquad i=1, \dots, N_f
\ee
with $s \to \infty$. This time there is a non-trivial SUSY vacuum at infinity in which, as we will see in details at the level of partition function in Section \ref{s3deriv}, half of the flavors remain massles and the gauge group is broken to $U(N_c)$ on the electric side, and to $U(N_f-N_c-2)$ on the magnetic side. The real mass deformation also reduces the number of massless singlets on the magnetic side to a $N_f^2$ block, organized in the matrix $M\ud{a}{b}$, which enters the magnetic superpotential as a Lagrange multiplier coupled to the dual mesons, $ \sum_{a,b=1}^{N_f} M\ud{a}{b} \tilde q_a q^b$.

The original $USp(2N_c)$ theory on $S^1$ had no topological nor axial symmetry. In the final $U(N_c)$ theory these symmetries are broken by non-perturbative effects, namely by the original instanton and by an extra non-perturbative Affleck-Harvey-Witten contribution \cite{Affleck:1982as} associated to the breaking of the gauge group $USp(2N_c)\to U(N_c)$. These two non-perturbative contributions can be identified with the sum of the two fundamental monopoles, $\CW_\text{mon}=\M^+ + \M^-$, breaking $U(1)_A \times U(1)_\text{T}$. The discussion on the magnetic side is similar. Eventually we arrive to the following duality between the electric theory $\CT_\M$ and a magnetic theory $\CT'_\M$: 
\be
\text{$\CT_\M :$} \qquad U(N_c) \text{ SQCD with $N_f$ flavors,} \quad \CW =\M^+ + \M^-
\ee
and
\bea
\text{$\CT'_\M :$} \qquad U(N_f - N_c - 2) &\text{ SQCD with $N_f$ flavors $q^i, \qt_i$ and $N_f^2$ singlets $M\ud{i}{j}$} , \\
&\CW = \sum_{i,j=1}^{N_f} M\ud{i}{j} \tilde q_i q^j+ \widehat\M^+ + \widehat\M^-  \;.
\eea
We discuss the map of the operators in the chiral ring in Section \ref{chiralringmap}. In Section \ref{reaha} we show, as a consistency check, that the duality $\CT_\M=\CT'_\M$ reduces to the Aharony duality for $U(N_c)$ theories \cite{Aharony:1997gp} after a suitable real mass deformation.  Further consistency checks via complex deformations are given in Section \ref{sec: complex mass def}.

An obvious question is whether we can also reach $\CT_\M$ starting from $\CT_0$, the fixed point of the 3D $\CN=2$ $U(N_c)$ gauge theory with $N_f$ chiral multiplets $Q^i$ in the fundamental and $\Qt_j$ in the antifundamental representation of $U(N_c)$ and $\CW=0$.
 The chiral ring of $\CT_0$ \cite{Aharony:1997bx, Aharony:1997gp} for $N_f \geq N_c-1$ is generated by a $N_f \times N_f$ matrix of mesonic fields $M\ud{i}{j} = Q^i \Qt_j$ and by two monopole operators $\M^+$, $\M^-$. In $\CT_0$ we can try to turn on the superpotential term $\CW_\text{\rm mon} = \M^+ + \M^-$ and reach $\CT_{\M}$.
As we will discuss in Section \ref{sec: RGflows}, for a  large portion of the range of $N_c, N_f$ the deformation $\CW_\text{\rm mon}$ is irrelevant at $\cT_0$ and therefore simply adding it to $\cT_0$ does not initiate an RG flow that leads to a new fixed point. 

However the non-trivial fixed point $\cT_\fM$ does exist for a larger window of parameters and the RG flow across dimensions we have just discussed 
provides a UV completion for $\CT_{\M}$. We postpone the question of how to reach the SCFT $\cT_\fM$ starting from a weakly coupled 3D Lagrangian model to Section \ref{sec: RGflows}. In the remaining of this section we will study $\cT_\fM$ for various ranges of values of $N_c$, $N_f$ without further inquiring how $\cT_\M$ is UV completed. UV completion for theories with monopoles in the superpotential have been discussed also in \cite{Intriligator:2013lca}.

Let us remark that, as long as the fixed point $\cal{T}_\M$ exists, we can study some of its properties---such as its moduli space or the anomalous dimensions of chiral operators---using $\mathcal{W}_\text{mon}$ even if such an operator is irrelevant at $\CT_0$.

\subsection[Basic properties of $\cT_\fM$]{Basic properties of $\boldsymbol{\cT_\fM}$}
\label{sec: basic properties}

In $\CT_{\M}$ the superpotential (\ref{Wmon}) has R-charge $2$, so the superconformal R-charge $R[Q]$ of the quarks $Q,\Qt$ can be computed imposing that the monopole operators, whose R-charge is given by the formula (\ref{monfor})
 \be
 R[\M^\pm] = N_f (1-R_Q) - N_c + 1 \;,
 \ee
 have R-charge $2$:
 \be
 \label{Rcharges}
 R[\M^\pm]_{\CT_{\M}} = 2 \quad\Rightarrow\quad R[Q]_{\CT_{\M}} = 1- \frac{N_c+1}{N_f} \;.
 \ee
 In $\CT_{\M}$ the operators $\M^\pm$ are not part of the chiral ring anymore: the deformation by the monopole superpotential $\CW_\text{mon}$ lifts the two branches of the Coulomb branch (when present) and the mixed branches parametrized by $\M^\pm$. By standard arguments \cite{Aharony:1997bx}, suppose to give a VEV to one of the monopoles $\M^\pm$. This breaks the gauge group to $SU(N_c-1) \times U(1)$ and at low energies the monopole operator becomes a fundamental field parameterizing the would be Coulomb branch, however the F-term potential following from $\CW_\text{mon} = \M^+ + \M^-$ provides a positive energy lifting those vacua. 
 
For $N_f \geq N_c+2$, the continuous flavor symmetry of the IR SCFT is $SU(N_f)^2$, and the $N_f^2$ generators of the chiral ring transform in the bifundamental representation. The pure Higgs branch is the space of $N_f \times N_f$ matrices $M$ with $\rank M \leq N_c$. This space has complex dimension $2N_c N_f-N_c^2$ and is generated by the $N_f^2$ mesons $M\ud{i}{j}$ subject to various non-independent relations:
\be
\label{rankN}
\epsilon_{i_1, \ldots, i_{N_f}} M\ud{i_1}{j_i} \ldots M\ud{i_{N_c+1}}{j_{N_c+1}} = 0 \;.
\ee
As we will see, for smaller values of $N_f$ the IR dynamics is different. In the $N_c=1$ we can use Abelian mirror symmetry \cite{Aharony:1997bx, Kapustin:1999ha} to check these statements, which we discuss in Appendix \ref{abeliansection}.

\subsection{Unitarity bound}

From equation (\ref{Rcharges}) we can find constraints on the previous discussion. The meson fields $M = Q\Qt$ must satisfy the unitarity bound:
\be
R[M]_{\CT_{\M}} = 2 R[Q]_{\cT_\fM} \geq \frac{1}{2} \qquad\Rightarrow\qquad  N_f \geq \frac43 (N_c+1) \;.
\ee
If $N_f > \frac43 (N_c+1)$ there can be an interacting SCFT in which the R-charges are as in (\ref{Rcharges}). If $N_f$ is smaller than or equal to the bound, the $N_f^2$ basic mesonic operators $M$ become free decoupled fields in the IR. This is somehow analogous to what happens in 4D $SU(N_c)$ SQCD with $N_f$ flavors when $N_f \leq \frac32 N_c$ \cite{Seiberg:1994pq}, or in 3D $\cN=4$ ``ugly'' or ``bad'' theories \cite{Gaiotto:2008ak}. Other examples with similar behavior have been studied in \cite{Kutasov:2003iy, Safdi:2012re}. One noteworthy aspect is that the theory breaks into a free sector and a leftover interacting SCFT. We will provide some evidence for this statement in Section \ref{sec: nfsmall}. In the dual SCFT, for $N_f \leq \frac43 N_c$ a particular cubic superpotential term, coupling the gauge-singlet mesons to the dual quarks, becomes irrelevant and must be dropped. This picture in the magnetic theory is consistent with having $N_f^2$ free mesons plus an interacting SCFT on the electric side.

\subsection{Map of the moduli space of vacua}
\label{chiralringmap}

As a first test of the duality $\cT_\M =\cT_\M'$  we show how the chiral ring generators are related.
In the magnetic theory $\cT_\M'$ imposing that the monopole superpotential has R-charge 2 allows us to extract the R-charge of the magnetic flavors:
\be
\label{R-charges mag}
R[\wh\M^\pm]_{\cT_\M'} = N_f \big( 1 - R(q) \big) - (N_f - N_c -2) + 1 = 2 \quad\Rightarrow\quad R[q]_{\cT_\M'} = \frac{N_c + 1}{N_f} \;.
\ee
The gauge-singlet fields $M\ud{i}{j}$ have thus R-charge 
\be
\label{mesonRcharges}
R[M\ud{i}{j}]_{\CT_\M'} = 2 \Big( 1- \frac{N_c+1}{N_f} \Big) \;,
\ee
matching the $N_f^2$ mesons $Q^i \wt Q_j$ of the electric theory (\ref{Rcharges}). These operators are the generators of the chiral ring and transform in the bifundamental representation of $SU(N_f)^2$.

In order to show that the moduli spaces of vacua match, we need to verify the relations satisfied by the generators. In the electric theory the relations are the ones in (\ref{rankN}) forcing the $N_f \times N_f$ matrix $Q^i \wt Q_j$ to have rank at most $N_c$.

In the magnetic theory, when the singlets $M\ud{i}{j}$ get a VEV of rank $r$, they give mass to $r$ of the $N_f$ flavors so the theory flows to $U(N_f-N_c-2)$ with $N_f-r$ flavors and same superpotential $\CW = M\ud{i}{j} \qt_i q^j + \wh \M^+ + \wh \M^-$ as before. If $r$ is larger than $N_c$, $\CT'_\M$ has no vacuum. To see that, we can perform Aharony duality \cite{Aharony:1997gp} leading to $U(N_c+2-r)$ with $N_f-r$ flavors and superpotential
\be
\CW = \fN^+ \wh S^- + \fN^- \wh S^+ + \wh S^+ + \wh S^- \;,
\ee
where $\wh S^\pm$ are now gauge singlets dual to the previous monopoles, while $\fN^\pm$ are the monopoles of the new description. The F-terms of $\wh S^\pm$ imply that both $\fN^+$ and $\fN^-$ must take a non-zero VEV, which in a $U(k)$ theory is compatible with supersymmetry only if $k>1$. We conclude that there are supersymmetric vacua only if $r \leq N_c$. The non-vanishing VEV of both $\fN^\pm$ breaks the gauge symmetry to $U(N_c -r)$, and if the singlets $M\ud{i}{j}$ have a VEV of maximal rank $r=N_c$ then there is no leftover gauge symmetry in the IR and one finds precisely one point in the moduli space.

We conclude that in both descriptions the moduli space of vacua is the set of $N_f \times N_f$ matrices with rank at most $N_c$.

\section{Dynamics of $\boldsymbol{\CT_\M}$ in the $\boldsymbol{(N_c, N_f)}$-space}
\label{Dynamics}

\begin{figure}[t]
\begin{center}
\includegraphics[height=.7\textwidth]{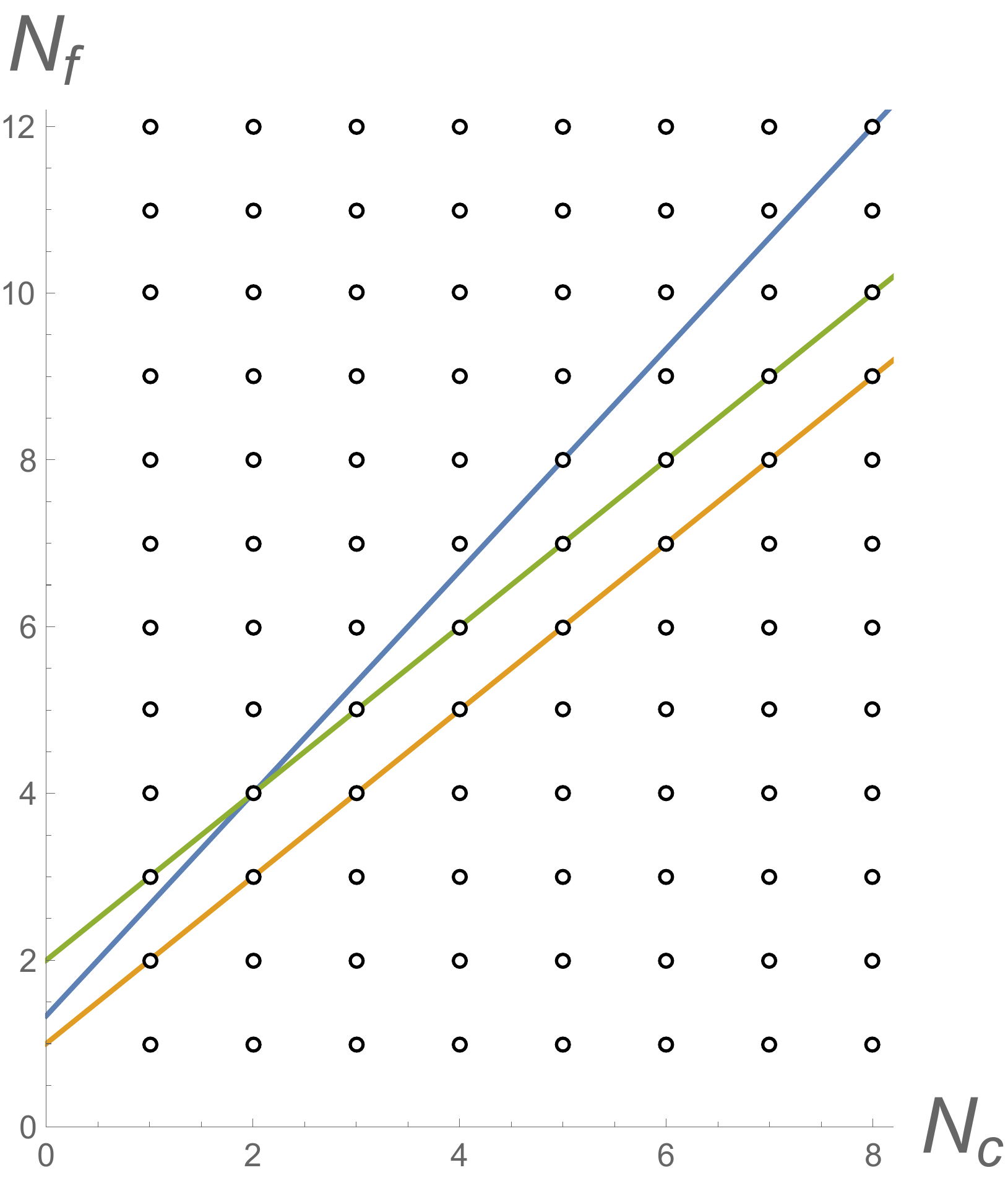}
\caption{Dynamics of $\cT_\fM$ in various regions of the parameter space $N_c, N_f$.
Blue line $N_f=\frac43(N_c+1)$, green line $N_f = N_c + 2$, orange line $N_f = N_c + 1$.
\label{fig: graph}}
\end{center}
\end{figure}

Our discussion of the Abelian case in Appendix \ref{abeliansection} and the observation on the constraints imposed by unitarity suggest that as we vary $N_c$, $N_f$  the theory has a non-trivial dynamics.  Indeed the picture is quite intricate and goes as represented in Fig.~\ref{fig: graph} and explained in the following.

\begin{itemize}

\item For $N_f \geq N_c + 3$, the IR limit $\cT_\fM$ contains an interacting factor and  has the dual description $\cT'_\fM$.
\begin{itemize}
\item For $N_f > \frac43(N_c+1)$ (above blue line) $\cT_\fM$ is completely interacting. 
\item For $N_c + 3 \leq N_f \leq \frac43 (N_c+1)$ (above green line and up to blue line) the IR limit $\cT_\fM$ breaks into an interacting SCFT and a decoupled free sector.
\end{itemize}

\item For $N_f = N_c + 2$ (green line), $\cT_\fM$ is described by a Wess-Zumino model. For $N_c=1$, $N_f=3$ this gives an interacting SCFT. For $N_c = 2$ the superpotential is marginally irrelevant and for $N_c \geq 3$ it is irrelevant, therefore $\cT_\fM$ is free.

\item For $N_f = N_c + 1$ (orange line) there is a smooth moduli space associated to the deformation, which is a quantum deformation of the classical Higgs branch, therefore the IR limit $\cT_\fM$ is free.

\item For $N_f \leq N_c$ (below the orange line) the theory has no supersymmetric vacua and $\cT_\fM$ does not exist.

\end{itemize}

\subsection[The region $N_c + 3 \leq N_f \leq \frac43(N_c+1)$: a decoupled sector]{The region $\boldsymbol{N_c + 3 \leq N_f \leq \frac43(N_c+1)}$: a decoupled sector}
\label{sec: nfsmall}

As we observed in Section \ref{sec: basic properties}, in the region $N_c + 3 \leq N_f \leq \frac43(N_c+1)$ if we use the R-charges in (\ref{Rcharges}) the mesons of the electric theory would violate the unitarity bound. We show here that in the dual theory $\CT'_{\M}$ in this region the superpotential terms $M\ud{i}{j} \qt_i q^j$ are actually irrelevant. Once we discard them, the $N_f^2$ gauge singlets $M\ud{i}{j}$ become free and decoupled. There are then accidental symmetries in the IR, such that there exist consistent R-charges that do not violate the unitarity bounds. This is what one would expect in the electric theory.

To see this, we start from a $U(N_c' =N_f-N_c-2)$ theory with $N_f$ flavors $q$, $\tilde q$ and monopole superpotential $\CW = \widehat \M^+ + \widehat \M^-$, that we can call $\cT$. We compute the R-charges of $q$, $\qt$ by setting $R[\widehat \M^\pm]=2$ and recalling that we are considering $N_f \leq \frac43(N_c+1)$ we find:
\be
R[Q]_\cT = 1 - \frac{N'_c+1}{N_f} = \frac{N_c+1}{N_f} \geq \frac34 \;,
\ee
hence in $\CT$ the mesons are above the  unitarity bound and the superpotential deformation $\CW_\text{def} = M\ud{i}{j} \tilde q_i q^j$ is irrelevant, $R[\CW_\text{def}] \geq 2$. We then propose that the dual description of $\cT_\fM$ for $N_c + 3 \leq N_f \leq \frac43 (N_c+1)$ is given by $N_f^2$ free chiral fields $M\ud{i}{j}$ together with a $U(N_f-N_c-2)$ gauge theory with $\CW= \widehat \M^+ + \widehat  \M^-$. For $N_f = \frac43(N_c+1)$ the superpotential deformation is marginally irrelevant.

Notice that $N_f \leq \frac43 (N_c+1)$ implies that the theory $U(N_c')$ has $N_f \geq 4 (N_c'+1)$ flavors. This is outside the range of parameters for which we have a 3D UV completion of $\CT_\M$ (see Section \ref{sec: RGflows}). We do not know whether, in this range of parameters, the superpotential $\CW = \widehat \M^+ + \widehat  \M^-$ leads to a fixed point distinct from the one for $\CW=0$ or not. In either case, the fixed point is interacting.

\subsection[The line $N_f = N_c+2$: a Wess-Zumino model]{The line $\boldsymbol{N_f = N_c+2}$: a Wess-Zumino model}
\label{WZM}

This case is more easily studied starting from the 4D duality for $USp(2N_c)$ SQCD \cite{Intriligator:1995ne} with $2N_f = 2N_c + 4$. 
The compactification to 3D and the real mass deformation (\ref{rmd}) produce an RG flow to the $U(N_c)$ theory with $N_f = N_c+2$ flavors and $\CW = \fM^+ + \fM^-$. In 4D, the dual magnetic theory is a Wess-Zumino model of $N_f (2N_f-1)$ gauge singlets $M_{ab}$ (organized into an antisymmetric matrix) with superpotential $\CW = \text{Pf}(M)$. The 3D compactification does not introduce non-perturbative effects, and after the real mass deformation only $N_f^2$ singlets $M\ud{i}{j}$ (organized into a $N_f \times N_f$ matrix) interacting with superpotential
\be
\CW = \det M
\ee
survive at low energies. The F-term equations following from the latter superpotential precisely impose the constraint that $\rank M \leq N_c$.

On the electric side, the requirement that  $R[\CW_\text{mon}]=2$ would fix $R[Q]=1/N_f$, therefore only the case $N_c=1$, $N_f=3$ satisfies the unitarity bound for the mesons. This case has been discussed at length in \cite{Benvenuti:2016wet}
(for $N_c=2$, $N_f =4$ the bound is saturated, and we expect the mesons to become free fields).  Correspondingly, on the magnetic side we find that the superpotential  $\CW = \det M$ is irrelevant for $N_f>4$ and marginally irrelevant for $N_f=4$. At the IR fixed point the massless degrees of freedom are $N_f^2$ free meson fields.

\subsection[Complex masses: consistency checks and the $N_f <N_c+2$ regions]{Complex masses: consistency checks and the $\boldsymbol{N_f <N_c+2}$ regions}
\label{sec: complex mass def}

We can perform simple consistency checks of the proposed dualities by taking complex mass deformations. We start in the region $N_f\geq N_c+3$ and consider a complex mass deformation of the electric side by the superpotential $\CW_\text{mass} = m Q^{N_f} \wt Q_{N_f}$. The total superpotential is thus
\be
\label{coma}
\CW_\text{el} = \fM^+ + \fM^- + m \,Q^{N_f} \wt Q_{N_f} \;.
\ee
In the IR we are left with $N_f-1$ flavors. Let us analyze the deformation in the magnetic $U(N_f - N_c - 2)$ description. The complex mass $\CW_\text{mass}$ is mapped to $m M\ud{N_f}{N_f}$, therefore the magnetic theory has superpotential
\be
\CW_\text{mag} = \sum_{i,j=1}^{N_f} M\ud{i}{j} \tilde q_i q^j + \wh\M^+ + \wh\M^- +  m M\ud{N_f}{N_f} \;.
\ee
By the F-term equations, the dual quarks get a VEV: $\tilde q_{N_f} q^{N_f} = -m$. Thus the gauge group is Higgsed to $U(N_f - N_c - 3)$, we are left with $N_f -1$ light flavors and a superpotential $\CW = \sum M\ud{i}{j} \tilde q_i q^j + \wh\M^+ + \wh\M^-$. This is consistent with the proposed duality.

If we start with $N_c+3$ flavors, the complex mass deformation takes the electric theory to the line $N_f = N_c+2$ flavors. On the magnetic side the $U(1)$ gauge group is completely Higgsed: we are left with $N_f^2$ chiral multiplets, and---because of complete Higgsing---instanton corrections produce the only superpotential compatible with the symmetries: $\CW=\det M$.

We can deform by a complex mass once more. On the electric side we add $m Q^{N_c+2} \tilde Q_{N_c+2}$, and flow to the theory with $N_f = N_c + 1$. On the magnetic side we have superpotential
\be
\CW= \det M + m M\ud{N_c+2}{N_c+2} \;.
\ee
The F-terms equations impose that $M\ud{i}{N_c+2} = M\ud{N_c + 2}{j} = 0$ for all $i,j$ and that
\be
\label{quantum deformed Higgs branch}
\det \wt M = - m
\ee
where $\wt M$ is the minor complementary to $M\ud{N_c+2}{N_c+2}$. Therefore, a dual description for $N_f = N_c+1$ is in terms of a non-linear sigma model of $N_f^2$ chiral superfields $\wt M\ud{i}{j}$ subject to the constraint (\ref{quantum deformed Higgs branch}). This could be described through a Lagrange multiplier $\lambda$ and a superpotential $\CW = \lambda( \det \wt M + m )$.

Finally we can add another complex mass to flow on the electric side to $U(N_c)$ with $N_f=N_c$ flavors. On the magnetic side, the IR dynamics is described by the superpotential
\be
\CW= \lambda \big( \det \wt M + m \big) + m \wt M\ud{N_c+1}{N_c+1} \;.
\ee
The resulting F-term equations do not have any solution and lead to runaway behavior.%
\footnote{Another way to see that for $N_f = N_c$ there is runaway behavior with no supersymmetric vacua is to use  the low energy description of $\cT_0$ \cite{Aharony:1997bx} as the Wess-Zumino model of $N_c^2 + 2$ chiral multiplets $\fM^\pm$, $M\ud{i}{j}$ with superpotential
\be
\label{Wess-Zumino description}
\CW = \fM^+ \fM^- \det M \;.
\ee
The addition of the monopole deformation $\CW_\text{mon} = \fM^+ + \fM^-$ leads to F-term equations with no solutions. The situation does not improve if we add masses for the flavors. For instance, to reach the case $N_f = N_c - 1$ we add a mass term for $Q^{N_f}$, $\Qt_{N_f}$. In the Wess-Zumino description (\ref{Wess-Zumino description}) this appears as a superpotential term $\CW_\text{mass} = M\ud{N_f}{N_f}$. Again the F-term equations do not have solutions.
}

\section{UV completions of $\boldsymbol{\CT_{\M}}$ in three dimensions}
\label{sec: RGflows}

As we have already mentioned, it is natural to wonder whether $\CT_\M$ can also be reached via more conventional three-dimensional RG flows in addition to the 4D $\to$ 3D flow discussed in Section \ref{4d3dflow}. We can think of starting with $U(N_c)$ SQCD with $N_f$ flavors, flow to its fixed point $\CT_0$, and then turn on the superpotential deformation $\CW_\text{mon}$ (\ref{Wmon}). In order to do that,  the deformation must be relevant and this happens for
\be
R[\M^\pm]_{\CT_0} = N_f \big( 1-R[Q]_{\CT_0} \big) -N_c + 1 < 2 \;,
\ee
where $R[Q]_{\CT_0}$  is the superconformal R-charge at the fixed point $\CT_0$ which depends on $N_c,N_f$. 
The numerical values of $R[\M]_{\CT_0}$ for small $N_c, N_f$ have been computed in \cite{Benini:2011mf,Safdi:2012re} and we report them in the Table (\ref{numertab}) for convenience:
\be \begin{array}{|c|ccccccc|}\hline
            &  N_f=1   &  N_f=2   &  N_f=3   &  N_f=4   &  N_f=5   &  N_f=6   &  N_f=7   \\ \hline
N_c=1 &     2/3      &   1.18     &  1.69        &    2.19      &    2.69     &   3.20    &   3.70  \\ 
N_c=2 &           &      1/2  &    0.97      &    1.46      &    1.95     &   2.44    &  2.94   \\ 
N_c=3 &           &           &          &   0.78       &   1.24      &    1.72   &   2.20  \\ 
N_c=4 &           &            &          &            &      0.60   &   1.03    &  1.50   \\  
N_c=5&            &            &          &             &              &            & 0.84\\  
\hline
\end{array}
\label{numertab}
\ee
We see that for each $N_c$, there are just few values for which  $1/2<R(\M)_{\CT_0}<2$.

For large $N_c$, $N_f$, \cite{Safdi:2012re} found that if $N_f \lesssim 1.45 \, N_c$, then $R[\M]_{\CT_0}<1/2$. In those cases the monopole operators become free decoupled fields in the IR and the superconformal R-symmetry is not visible in the UV description: it mixes with an IR accidental symmetry. Then we cannot deform the IR fixed point, because the deformation $\CW_\text{mon}$ by free fields would break supersymmetry. Moreover, at the bound $N_f \approx 1.45\, N_c$ the R-charge of the fundamental fields is $R[Q]_{\CT_0} \approx 1-1/1.45 \approx 0.31$. If we call $N_{f,0}$ the smallest value of $N_f$ (at fixed $N_c$) for which $R[\fM^\pm]_{\cT_0} > \frac12$, then as we increase $N_f$ by a single unit beyond $N_{f,0}$ we find that $R[\M]_{\cT_0}$ increases approximately by $1-R[Q]_{\cT_0} \approx 0.69$. We conclude that for large fixed $N_c$ there are only two or three values of $N_f$ for which $\frac12 < R[\fM^\pm]_{\cT_0} < 2$ and $\CW_\text{mon}$ is a good relevant deformation.

However we can try to start from $\cT_0$ plus some other decoupled sector, couple them together and trigger a non-trivial RG flow. 

At this point we can stop to make a (partial) analogy with the case of the non-supersymmetric Gross-Neveu model with $N$ fermions, described by the Lagrangian $\bar{\Psi}_I \partial \Psi^I + (\bar{\Psi}_I \Psi^I)^2$. This theory has a ``UV fixed point'', which cannot be reached from the CFT of $N$ free fermions since the term $(\bar{\Psi}_I \Psi^I)^2$ has scaling dimension $4$ and is irrelevant in the free CFT. However, one can start from $N$ free fermions plus an Ising-CFT $\CT_{\sigma^4}$, that is a real scalar $\sigma$ with $\sigma^4$ potential at the Wilson-Fisher fixed point, and turn on the relevant deformation $\sigma \bar{\Psi}_I \Psi^I$. This is called the Gross-Neveu-Yukawa model. In the infrared it flows to a fixed point, that we call Gross-Neveu$_N$. Such a fixed point can be further deformed by $\sigma^2$, and the resulting RG flow connects to the Gross-Neveu model (which describes the leading irrelevant operator along the flow). In this sense, Gross-Neveu$_N$ can be though of as the UV fixed point we were after. We can represent the RG flow as
\be
N \text{ free fermions } \oplus \CT_{\sigma^4} \,\,
\underrightarrow{\qquad \text{Gross-Neveu-Yukawa}\qquad} \,\,
\text{Gross-Neveu}_N \;.
\ee

The difference with our case of $\CT_0$ and $\CT_\M$ is that, as we explained, there exists an RG flow Gross-Neveu$_N$ $\to$ $N$ free fermions, while in our case we do not expect an RG flow from $\CT_\M$ to $\CT_0$ (for the same values of $N_f, N_c$). One way to show that would be to verify that $\CF_{\CT_\M}[N_f,N_c] < \CF_{\CT_0}[N_f,N_c]$ (where $\cF$ is the Euclidean free energy on $S^3$ \cite{Jafferis:2010un}) and hence an RG flow from $\CT_\M$ to $\CT_0$ would violate the $\CF$-theorem \cite{Jafferis:2011zi, Casini:2012ei}. The inequality $\CF_{\CT_\M}[N_f,N_c] < \CF_{\CT_0}[N_f,N_c]$ follows if one proves that $\CF[N_f,N_c](r_Q)$, where $r_Q$ is the R-charge of the quarks, is concave with a maximum (which corresponds then to $\CT_0$) in the physically sensible interval $\frac14 \leq r_Q \leq 1$. This can be numerically checked for small values of $N_f, N_c$.

\begin{figure}[t]
\begin{center}
\includegraphics[width=.4\textwidth]{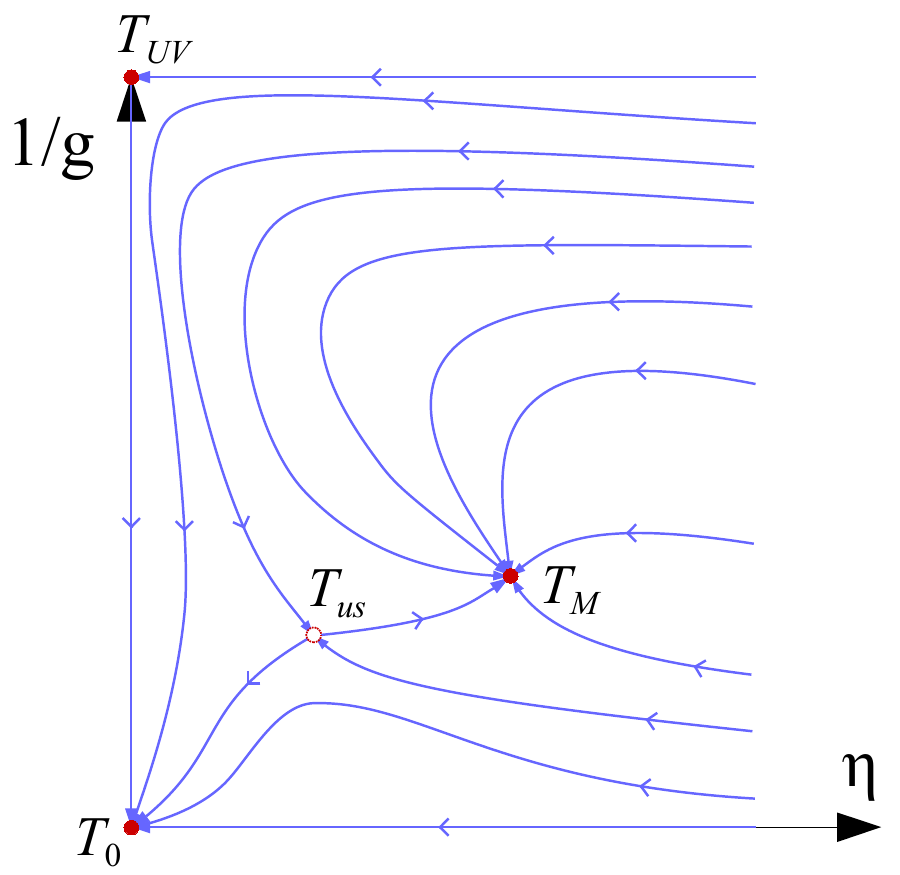}
\caption{Qualitative RG diagram of the supposed minimal flow that can accommodate $\CT_0$ and $\CT_\M$ in cases that $\CW_\text{mon}$ is irrelevant in $\CT_0$. Here $g$ is the gauge coupling and $\eta$ the monopole coupling in $\eta \CW_\text{mon}$. The point $\CT_\text{UV}$ is the weakly-coupled $U(N_c)$ SQCD with $N_f$ flavors, $\CT_0$ is its IR fixed point, and $\CT_\M$ the non-trivial fixed point with monopole deformation turned on. The topology requires the existence of (at least) one unstable fixed point $\CT_\text{us}$.
\label{figRGdiag}}
\end{center}
\end{figure}

We cannot tell what is the exact RG diagram of $U(N_c)$ SQCD with gauge coupling $g$ and deformation $\eta \CW_\text{mon}$, but we can draw the qualitative structure of the minimal topology that accommodates the features we have described, in the cases that $\CW_\text{mon}$ is irrelevant in $\CT_0$ and yet $\CT_\M$ exists. This is depicted in Fig.~\ref{figRGdiag}.

\

For completeness, let us mention here that there is actually an RG flow  
\be \CT_\M [ N_f+2,N_c ] \longrightarrow \CT_0 [N_f, N_c] \,.\ee
The idea is to start from the $\CT_\M=\CT'_\M$ pair with $N_f+2$ flavors and consider  the real mass deformation: 
\ben
\nn&&m_{N_f+1}\to  m_{N_f+1}  +t\,, \qquad  \tilde m_{N_f+1}\to  m_{N_f+1} -t\\
&&m_{N_f+2}\to  m_{N_f+2}  -t\,, \qquad  \tilde m_{N_f+2}\to  m_{N_f+2} +t\,,
\een
with $t\to\infty$.  As we explicitly show at the level of the partition function in Section \ref{reaha}, the limit restores the 
$U(1)_{A}\times U(1)_{T}$ and on the electric side we recover $\CT_0$.
On the dual side the  limit also reduces the $(N_f+2)\times (N_f+2)$ singlets to an $N_f\times N_f$ block which enters the superpotential in a cubic coupling with  the dual quarks plus  two extra singlets  coupling linearly to the dual monopoles. So on the dual side we recover $\CT'_0$ with   $\CW=\sum_{a,b}^{N_f}   M\ud{a}{b} \tilde q_a q^b +\widehat \M^+ S^-+ \widehat\M^- S^+$.

\subsection{UV completion in 3D using auxiliary Ising-SCFTs}
\label{superisi}

Going back to the supersymmetric case, let us add to $\cT_0$ $N_f^2$ copies of the Ising-SCFT, each consisting of a chiral superfield $\Phi_{ij}$ ($i,j=1,\cdots N_f$) with a cubic superpotential $\CW=\Phi_{ij}^3$ fixing $R[\Phi_{ij}]=2/3$. We then turn on a cubic superpotential to couple the singlets $\Phi_{ij}$ to the quarks in $\CT_0$:
\be
\CW= \sum_{ij=1}^{N_f} \left( \Phi_{ij} Q_i \Qt_j + \Phi_{i j}^3 \right) \;.
\ee
This deformation is relevant since $R[\Phi_{ij}]=2/3$, $R[Q]_{\CT_0}<1/2$ and breaks the continuous global symmetries $SU(N_f)^2 \times U(1)_A\rightarrow S_{N_f}\times S_{N_f}$, leaving enough discrete symmetries to set all the R-charges of the quarks equal to each other. This cubic coupling is expected to trigger an RG flow to the theory $\cT_{\Phi_{i j} Q_i \Qt_j+\Phi_{ij}^3}$ with $R[Q]=R[\Phi_{ij}]=2/3$ and
\be
R[\M^\pm] = N_f \big(1-R[Q] \big)-N_c +1=  N_f \big(1-\tfrac{2}{3} \big)-N_c +1 \;.
\ee

More precisely, there are two options for what the flow is. The first one, described above, leads to the theory $\cT_{\Phi_{i j} Q_i \Qt_j+\Phi_{ij}^3}$ with $R[Q] = R[\Phi_{ij}] = 2/3$. The other flow leads to a theory $\cT_{\Phi_{i j} Q_i \Qt_j}$ with $R[\Phi_{ij}]>2/3$. We later discuss that if we do not consider enough copies of the Ising-SCFT, the first flow can violate the $\cF$-theorem, hence the theory must follow the second flow. With enough $\Phi_{ij}$ it might be possible to use similar arguments to rule out the second flow, but we will not do it here.

The monopoles do not violate the unitarity bound, \ie{} they have $R[\M^\pm]>\frac12$, if $N_f \geq 3N_c-1$.
If we set $N_f= 3 N_c +2$ we find
\be
\label{rm1ststep}
R[\M^\pm] = N_f \big(1-R[Q] \big)-N_c +1= \frac{5}{3} < 2 \;,
\ee
so the monopoles in this case are a relevant deformation and we can add  $\M^+ + \M^-$ to the superpotential. This further deformation takes us to a theory with  $R[Q]_{\cT}=\frac{2N_c+1}{3N_c+2}$ and $R[\Phi_{ij}]=2-2R[Q]=\frac{2N_c+2}{3N_c+2}>2/3$, so the $N_f^2$ terms $\Phi_{ij}^3$ should actually be dropped.
In order to reach $\cT_\M$ we only need to get rid of the $N_f^2$ singlet fields $\Phi_{ij}$. To do so we add $N_f^2$ free chiral fields $\sigma_{ij}$ and couple them linearly to the $\Phi_{ij}$'s. Both $\Phi_{ij}$ and $\sigma_{ij}$ become massive and integrating them out we finally flow to $\cT_{\M}$.

Summarizing, for $N_f=3N_c+2$ we have the chain of 3D unitary RG flows
\be
\label{chainRGf}
\CT_0 \oplus \cT_{\Phi_{i j}^3} \oplus \sigma_{i j} \;\rightarrow\; \CT_{\Phi_{i j} Q_i \Qt_j + \Phi_{i j}^3} \oplus \sigma_{i j} \;\rightarrow\; \CT_{\M+\Phi_{i j} Q_i\Qt_j} \oplus \sigma_{i j} \;\rightarrow\; \CT_{\M} \;.
\ee
We can reach theories with less flavors $N_f<3N_c+2$ by simply adding complex mass terms for the quarks.%
\footnote{ One could also try to start with the Ising SCFT theory with $N+1$ chirals and superpotential $\CW= \Phi_{N+1} \sum_{i=1}^N \phi_i^2.$
At large $N$ the R-charge of the chirals become $R[\phi_i]\to 1/2$ and  $R[\Phi_{N+1}]\to 1$ respectively. We could then couple the mesons of the $\CT_0$ theory to the singlets $\phi_i$ by the cubic superpotential $Q\tilde Q (\sum_i^N \phi_i)$. At large $N$ this term would drag the R-charge of the quarks to $R[Q]\to 3/4$, which in turn would imply that $\CW_{\rm mon}$  remains a relevant deformation  for $N_f<4(N_c+1)$. One would need to check that there are no $\CF$-theorem violations along this flow as we discuss in the case of single Ising SCFT.}

\subsubsection*{The case $\boldsymbol{N_f=3(N_c+1)}$: conformal manifold}

 We could also start from $N_f=3(N_c+1)$. In this case, after coupling $\CT_0$ to the Ising-SCFTs, $R[\M^\pm]=2$ so there are two marginal monopole couplings while there are no mesonic operators of $R$-charge two. Since the two marginal deformations break only one $U(1)$ global symmetry (the topological symmetry, while the axial symmetry is already broken) there is precisely one exactly marginal monopole direction \cite{Kol:2002zt, Benvenuti:2005wi, Green:2010da, Kol:2010ub}. It is natural to parameterize this  exactly marginal deformation by $\M^++\M^-$, preserving the charge-conjugation $\IZ_2$ symmetry and the $S_{N_f} \times S_{N_f}$ permutation symmetry of the $N_f$ quarks. Our proposal is that in this $1$-complex dimensional%
 \footnote{The dimension of the full conformal manifold is bigger, since we can turn on many cubic couplings of the form $\Phi_{ij}\Phi_{kl}\Phi_{mn}$ and $\Phi_{ij}Q_k\Qt_l$, breaking the discrete symmetries.}
conformal manifold there is a point corresponding to $\CT_{\M}$, where the couplings with Ising-SCFTs are tuned to zero and the Ising-SCFTs decouple.

There is an analog situation in $U(1)$ with $N_f=1$ flavors where the three operators $\M^3$, $\M^{-3}$ and $(Q\Qt)^3$ have $R=2$. Since these operators break the $U(1)_T \times U(1)_A$ symmetry, there is one exactly marginal direction given by $\M^3 + \M^{-3} + (Q\Qt)^3$, preserving the $\IZ_3$ symmetry which is evident in the dual $XYZ$ description where the superpotential becomes:
\be
\CW = \lambda_1 X Y Z + \lambda_2 ( X^3 + Y^3 + Z^3 ) \;,
\ee
with $\lambda_{1,2}$ parameterizing a $\IC\IP^1$. At the point $\lambda_1=0$ the theory factorizes into three copies of the Ising-SCFT. This is the analog of the point  $\CT_{\M}$, where the couplings with Ising-SCFT's are turned off.

\subsection*{The reason for using $\boldsymbol{N_f^2}$ Ising-SCFTs}

The reader might wonder why we used $N_f^2$ auxiliary fields $\Phi_{ij}$ instead of just one $\Phi$. The reason is as follows.
Imagine that by coupling the $N_f^2$ mesons to a single Ising-SCFT we  reach the theory $\CT_{\Phi Q\Qt + \Phi^3}$ 
where $R[Q]=R[\Phi]=2/3$. At this point we could  couple  $\Phi$ to a single chiral field $\sigma$, integrate $\Phi$ and $\sigma$ out and flow  to $\CT_{0}$. We claim that this flow can violate the $\CF$-theorem \cite{Jafferis:2011zi, Casini:2012ei}  (for a review see \cite{Pufu:2016zxm}) for $N_f$ and/or $N_c$ large enough.
 
Let $\cF_{SQCD}[r]=-\log|Z_{SQCD}|$ be the $S^3$ free energy of the $U(N_c)$ SQCD with $N_f$ flavors as a function of the R-charge of the quarks $r$.  $\cF_{SQCD}[r]$ can be computed via localization \cite{Kapustin:2009kz, Jafferis:2010un, Hama:2010av} and according to $\CF$-maximization \cite{Jafferis:2010un} it is (locally) maximized at the value $r^*$ which is the  IR superconformal R-charge. In particular the free energy of a chiral multiplet of R-charge $r$ is given by the function $-\ell[1-r]$ defined in  \cite{Jafferis:2010un}, which is (locally) maximized at the value $r^*=1/2$ corresponding to that of a free 3D chiral field.

Now, consider our hypotetical flow from  $\CT_{\Phi Q\Qt + \Phi^3}\oplus \sigma$ in the UV to $\CT_0$ in the IR. $\cF_{UV}= \cF[\CT_{\Phi Q\Qt + \Phi^3}\oplus \sigma]
= \cF_{SQCD}[\tfrac{2}{3}]-\ell[1-\tfrac{2}{3}]-\ell[1-\tfrac{1}{2}]$ (the last two contributions are due to the chiral $\Phi$ and the free chiral $\sigma$), while $\cF_{IR}= \cF_{SQCD}[r^*]$, so
\be
\cF_{UV} - \cF_{IR} = \cF_{SQCD}[\tfrac{2}{3}] - \cF_{SQCD}[r^*] -\ell[1-\tfrac{2}{3}]-\ell[1-\tfrac{1}{2}] \;
\ee
The contribution  $\cF_{SQCD}[\tfrac{2}{3}] - \cF_{SQCD}[r^*]$ grows in modulus with $N_f,N_c$ and is negative (since $\cF_{SQCD}[r]$ is locally maximized at $r=r^*$ and assuming we are not too far away from this local maximum), while the contribution $-\ell[1-\tfrac{2}{3}]-\ell[1-\tfrac{1}{2}]$ from the two singlets is independent of $N_f,N_c$. Hence, for $N_f,N_c$ large enough, $\cF_{UV} - \cF_{IR}$ is negative and the $\CF$-theorem is violated.%
\footnote{For example, for $N_c=2, N_f=7$, one finds $\CF[r^*]=12.38$ and $\ell[\tfrac{1}{2}]=-0.347$ while $\CF[\tfrac{2}{3}]=8.75$,  $\ell[1-\tfrac{2}{3}]=-0.291$, so  $\cF_{UV}=8.75+0.291+0.347< \cF_{IR}=12.38$.}

 The conclusion is that $\CT_{\Phi Q\Qt + \Phi^3}$ does not exists with only one singlet $\Phi$, more precisely the $\Phi^3$ becomes irrelevant and must be dropped when coupling the Ising-SCFT $\Phi^3$ to the mesons of the SQCD. See \cite{Chester:2015qca} for an analogous discussion in a cubic Wess-Zumino model with $N+1$ fields. This problem however can be avoided  if we add $N_f^2$ copies of the Ising-SCFT $\Phi_{i j}$.

\subsection{Dual 3D RG flows}
\label{dualrg}

In this section we study the dual of the chain of RG flows (\ref{chainRGf}) to arrive at the dual theory $\CT'_{\M}.$ We start from the Aharony dual of $U(N_c)$ with $N_f$ flavors, add $N_f^2$ copies of Ising-SCFTs $\Phi_{i j}$, and couple the $\Phi_{i j}$ singlets to the dual mesons $M_{i j}$, which are themselves gauge singlets:
\be
\CW = \sum_{i,j=1}^{N_f} \big( M_{i j} \qt_i q_j  + M_{i j} \Phi_{i j} + \Phi_{i j}^3 \big) + S^+  \widehat\M^++ S^-  \widehat\M^- \;.
\ee
All the $M_{i j}$ and $\Phi_{i j}$ become massive and integrating them out we are left with a sextic superpotential in the quarks:
\be
\CW =- \sum_{i,j=1}^{N_f} (q_i \qt_j)^3 + S^+  \widehat\M^++ S^-  \widehat\M^- \;,
\ee
which sets $R[q]=\frac13$, $R[ \widehat\M^\pm]=-\frac13 N_f + N_c + 1$ and $R[S^\pm]=\frac13 N_f - N_c + 1$. It follows $R[S^\pm]>\frac12$ so unitarity is not violated, as long as $N_f \geq 3 N_c -1$. The singlets $\Phi_{i j}$ in the electric theory $\CT_{\Phi_{i j} Q_i\Qt_j + \Phi_{i j}^3}$ are mapped to the mesons $q_i \qt_j$ in the magnetic theory. The mesonic chiral ring is truncated (in the magnetic side this is due to the F-terms of $q_i$, $\qt_j$) in both dual theories.

At this point in the electric side we set $N_f=3N_c+2$ and turned on the deformation $\CW_\text{mon} =  \M^+ + \M^-$, flowing to $\CT_{\M+\Phi_{i j} Q_i\Qt_j}$. In the magnetic side this corresponds to turning on $\CW=S^+ + S^-$. This superpotential pushes down to zero $R[ \widehat\M^\pm]$ and pushes up $R[q]$, making the sextic superpotential $(q \qt)^3$ irrelevant, so we must drop the sextic superpotential. Integrating $S^\pm$ out breaks the magnetic gauge group according to $U(N_f-N_c) \to U(N_f-N_c-2)$, which generates a monopole superpotential $ \widehat\M^+ +  \widehat\M^-$ in the IR:
\be
\CW=  \widehat\M^+ +  \widehat\M^- \;.
\ee
The last step is to linearly couple the $\Phi_{ij}$ in the electric theory to the $N_f^2$ singlets $\sigma_{ij}$. In the magnetic theory this corresponds to coupling the mesons $q_i \qt_j$ to the  singlets $\sigma_{ij}$, ending up precisely with $\CT'_\M$:
\be
\CW = \sum_{i,j=1}^{N_f} \sigma\ud{i}{j} \qt_i q^j+ \widehat\M^+ + \widehat\M^- \;.
\ee
Summarizing, it is possible to follow the chain of RG flows (\ref{chainRGf}) also in the dual description. We interpret the self-consistency of this picture as a strong hint that the 3D UV completion is correct and works as described.

\subsection[RG flows from $\cT_{\M}$]{RG flows from $\boldsymbol{\cT_{\M}}$}

We can also explore the possibility of flowing  away from $\CT_\M$ without changing $N_f,N_c$. In $\CT_\M$  when the meson has R-charge less than $3/2$, i.e. when $N_f < 4(N_c+1)$, we can turn on the coupling $\sigma_{ij }Q_i \Qt_j$  to $N_f^2$ free chirals $\sigma_{ij}$. At the end of this flow $R[\sigma_{ij}]=2\frac{N_c+1}{N_f}$. Now if $N_f>2(N_c+1)$, $\sigma_{ij}^2$ are relevant deformations, and turning them on fixes $R[\sigma_{ij}]=1$, $R[Q]=1/2$ and $R[\M^\pm]=N_f/2-N_c+1>2$. So the monopole superpotential has R-charge greater than $2$ and must be dropped and we are left with $\cT_\text{quartic}$.
 
Summarizing, for $2(N_c+1) < N_f < 4( N_c+1)$ we have the following chain of 3D unitary RG flows:
\be
\CT_\M \oplus \sigma_{ij} \;\longrightarrow\; \CT_{\M + \sigma_{ij} Q_i\Qt_j} \;\longrightarrow\; \CT_{\sigma_{ij} Q_i\Qt_j +\sigma^2_{ij}} = \cT_\text{quartic} \;.
\ee
If $3(N_c+1)<N_f<4(N_c+1)$, similar arguments show the existence of the RG flows
\be
\CT_\M \oplus \sigma_{ij} \;\longrightarrow\; \CT_{\M + \sigma_{ij} Q_i\Qt_j} \;\longrightarrow\; \CT_{\sigma_{ij} Q_i\Qt_j +\sigma^3_{ij}} \;.
\ee
Notice that, for a given $N_f$ and $N_c$, using the RG flows discussed in this section it is not possible to go back and forth from $\cT_\M$ and $\cT_\text{quartic}$ or $\CT_{\sigma_{ij} Q_i\Qt_j +\sigma^3_{ij}}$, even adding free chiral fields or $\Phi^3_{ij}$ SCFTs. In the cases $N_f=2(N_c+1)$ or $N_f=3(N_c+1)$ there are non-trivial conformal manifolds and we can continuously turn on and off the monopole superpotential.

\section{$\boldsymbol{\CT_\M}$ as the S-duality wall for 4D $\boldsymbol{\CN=2}$ SQCD}
\label{sec:dualitywall}

In this section we will see how the theory $\CT_\M$ with $N_f=2N_c+2$ can be identified with the S-duality wall for the  4D $\CN=2$ SQCD.
In the context of the AGT  correspondence relating Toda correlators to $S^4$ partition functions of class-$S$ theories  \cite{Alday:2009aq}, 3D S-duality walls were conjectured to be mapped to the elements of the Moore-Seiberg  groupoid  \cite{Moore:1988uz} acting on the conformal blocks \cite{Drukker:2010jp}.
In particular, a 3D interface theory can be placed on the three-sphere at the equator of the $S^4$, separating the two hemispheres where the 4D theories have coupling  related by a generalized  $S$-duality and the $S^3_b$ partition function of the interface theory is conjectured to be equal to the CFT kernel implementing the action of the Moore-Seiberg groupoid element.

The duality kernels in Liouville theory were obtained in \cite{Ponsot:1999uf,Ponsot:2000mt}. The $S$-kernel was shown to perfectly match with the $S^3_b$ partition function of the mass deformed $TSU(2)$ theory \cite{Hosomichi:2010vh}. The interpretation of the Liouville $F$-kernel as a domain wall theory, instead, has created some troubles. The matter content was immediately identified ($U(1)$ with $N_f=4$ flavors) but the identification required to impose various constraints on the real mass parameters and the origin of these constraints was not explained. Indeed, using various integral identities, in \cite{Teschner:2012em} the Liouville $F$-kernel was rewritten in a form that could be mapped to an $SU(2)$ partition function with no mass constraints.

In Toda CFT the braiding kernel for the 4-point block with two semi-degenerate vertex operators has recently been derived by Le Floch in \cite{Floch:2015hwo}. From the explicit form of the kernel it is easy to read out the matter content of the interface theory which was identified as a $U(N-1)$ theory with $2N$ chirals of charge $1$ and $2N$ chiral of charge $-1$. This theory is supposed to be self-dual and this property at the level of partition function follows from an integral identity. However, as in the Liouville case, the identification requires to impose a constraint on the real mass parameters associated to the topological and axial symmetries $U(1)_{T} \times U(1)_{A}$. It is also necessary to fix the R-charges of the quarks $R[q]$ to $\frac12$,  which \cite{Floch:2015hwo} conjectured to follow by the cubic coupling of the 3D quarks to the 4D hypers.

Here we point out that the 3D S-duality wall theory is actually $\CT_\M$, the theory analyzed in this paper, in the case $N_f=2(N_c+1)$. The monopole superpotential $\CW_\text{mon}= \M^+ + \M^-$ sets $R[q]=\frac12$ with no need of the cubic coupling to the 4D hypers and it breaks the $U(1)_{T}\times U(1)_{A}$ symmetries, implying that the corresponding real mass deformations cannot be turned on. The self-duality property of this theory is then a particular case of our $\CT_\M=\CT'_\M$ duality.

 
 \section{Higher monopole superpotentials}
 \label{sec: higher powers}
 
It is also interesting to consider superpotentials containing non-minimal chiral monopole operators. In the non-SUSY case,  potentials of the type $\M^k+(\M^{\dagger})^{k}$ with $k=2,3,4$, where $\M$ is now the minimal monopole, can arise in the thermodynamic limit of spin-models on lattices with $\IZ_k$ rotational symmetries (i.e. rectangular, honeycomb and square lattices), see for instance \cite{block2013fate}.
 
In this section we briefly discuss 3D UV completions of the  $U(N_c)$ SQCD with a superpotential quadratic or cubic in the basic monopole operators: $\CW=\M^{+2}+\M^{-2}$ and $\CW=\M^{+3}+\M^{-3}$.  We leave a more exhaustive analysis of the dynamics of these theories, including the discussion of their chiral rings and study of their potential derivation from 4D,  for future work.

\subsection[$\CW=\M^{+2}+\M^{-2}$]{$\boldsymbol{\CW=\M^{+2}+\M^{-2}}$}

Let us denote by $\CT_{\M^2}$ the SQCD with monopole superpotential $\CW=\M^{+2}+\M^{-2}$. Imposing  the marginality of the superpotential  $R[\CW]=2$ we get
\be
2R[\M^\pm]=2 \qquad\Rightarrow\qquad  R[Q]=\frac{N_f-N_c}{N_f} \;.
\ee
The unitarity bound for the meson is satisfied if 
\be
2R[Q]>\frac12 \qquad\Rightarrow\qquad N_f>\frac43N_c \;.
\ee
Proceeding as in Section \ref{superisi} we can find a 3D UV completion. We couple $\CT_0$ to $N_f^2$ copies of the Ising-SCFT through the cubic coupling, which sets $R[Q]=\frac23$. We take $N_f=3N_c-1$ which (with $R[Q]=\frac23$) gives $R[\M^{\pm}]=\frac23$. The superpotential deformation $\CW=\M^{+2}+\M^{-2}$ is relevant and can be turned on. This takes us to a theory with  $R[Q] = \frac{2N_c-1}{3N_c-1}$ and $R[\Phi_{ij}] = \frac{2N_c}{3N_c-1}>\frac23$, so the $N_f^2$ terms $\Phi_{ij}^3$ should actually be dropped from the superpotential. Finally, to reach $\cT_{\M^2}$ we get rid of the $N_f^2$ singlet fields $\Phi_{ij}$ by  coupling them linearly to  $N_f^2$ free chiral fields $\sigma_{ij}$. Both $\Phi_{ij}$ and $\sigma_{ij}$ become massive and integrating them out we finally flow to $\cT_{\M^2}$.
Giving masses to some quarks, we have a 3D UV completion for $\CT_{\M^2}$ for all $\frac43N_c < N_f < 3N_c$. 

We propose that an Aharony-Seiberg duality for $\CT_{\M^2}$ works as follows:

\be
\text{$\CT_{\M^2} :$} \qquad U(N_c) \text{ SQCD with $N_f$ flavors,} \quad \CW =(\M^+)^2 + (\M^-)^2
\ee
and
\bea
\text{$\CT'_{\M^2} :$} \qquad U(N_f - N_c) &\text{ SQCD with $N_f$ flavors $q^i, \qt_i$ and $N_f^2$ singlets $M\ud{i}{j}$} , \\
&\CW = \sum_{i,j=1}^{N_f} M\ud{i}{j} \tilde q_i q^j+( \widehat\M^+)^2 + (\widehat\M^-)^2  \;.
\eea
To arrive at this duality one can start from the Aharony duality and turn on the quadratic monopole superpotential on the electric side. On the magnetic side this amounts to turning on the superpotential $(S^+)^2+(S^-)^2$ where the singlets $S^\pm$ enter the superpotential 
as $\widehat\M^+ S^-+ \widehat\M^- S^+ $. Using the equations of motion for the two singlets $S^\pm$ we obtain the quadratic terms $\widehat\M^{+2}+\widehat\M^{-2}$ in the monopoles of the magnetic theory.

As a consistency check we can study the dual of the RG flow as in Section \ref{dualrg} to arrive to the dual theory $\CT'_{\M^2}$.
Adding the Ising-SCFTs to the Aharony dual of $U(N_c)$ with $N_f$ flavors and coupling the singlets $\Phi_{i j}$ to the dual mesons $M_{i j}$ we obtain the  theory with sextic superpotential in the quarks. In this theory the R-charges are set to $R[q]=\frac13$, $R[ \widehat\M^\pm]=-\frac13 N_f + N_c + 1$ and $R[S^\pm]=\frac13 N_f - N_c + 1$. 
Paralleling the steps on the electric side we set $N_f=3N_c-1$ and turn on the dual of the quadratic monopole deformation  $\CW=(S^+)^2 + (S^-)^2$. This superpotential pushes up the quark $R$-charges ($R[q]>1/3$) making the sextic superpotential $(q \qt)^3$ irrelevant. Integrating $S^\pm$ out generates the monopole superpotential $ (\widehat\M^+)^2 +  (\widehat\M^-)^2$ in the IR. The last step is to linearly couple the singlets $\Phi_{ij}$ in the electric theory to the $N_f^2$ singlets $\sigma_{ij}$. On the magnetic side this corresponds to turning the coupling $\sigma_{ij} q_i \qt_j$, hence we end up precisely with $\CT'_{\M^2}$.

\subsection[$\CW=\M^{+3}+\M^{-3}$]{$\boldsymbol{\CW=\M^{+3}+\M^{-3}}$}

We close this section with a brief discussion of the cubic monopole superpotential.  Let us denote by $\CT_{\M^3}$ the SQCD with monopole superpotential $\CW=\M^{+3}+\M^{-3}$. Imposing  the marginality of the superpotential  $R[\CW]=2$ we get
\be
R[\M^\pm]=\frac23 \qquad  \Rightarrow  \qquad  R[Q]=1-\frac{N_c-\frac13}{N_f} \;,
\ee
and the unitarity bound for the meson is satisfied when
\be
2R[Q]>\frac12 \qquad\Rightarrow\qquad N_f>\frac43 \Big( N_c-\frac13 \Big) \;.
\ee
As in quadratic monopole case, we can find a 3D UV completion by coupling $\CT_0$ to $N_f^2$ Ising-SCFTs. For $N_f=3N_c-1$ there is a conformal manifold where the monopole superpotential $\CW=\M^{+3}+\M^{-3}$ is exactly marginal. Giving masses to some quarks, we have a 3D UV completion for $\CT_{\M^3}$ for all $\frac43 (N_c-\frac13) < N_f < 3N_c$.


\section{$\boldsymbol{S^3_b}$ partition functions: dualities as integral identities}
\label{partfun}

In this section we check the duality $\CT_\M=\CT'_\M$ as well as consider various real mass deformations, at the level of the squashed three-sphere partition function $Z_{S^3_b}$ which can be computed via SUSY localisation as shown in  \cite{Kapustin:2009kz, Jafferis:2010un, Hama:2011ea} (for a review see \cite{Pestun:2016zxk}). Each chiral multiplet of  R-charge $r$ and real mass $m$ for its $U(1)$ flavor symmetry, contributes to the partition function as
\be
Z_\text{chiral}=s_b \big(\tfrac{iQ}{2}(1-r)-m \big)\quad {\rm with}\quad s_b(x) = \prod_{l,n {\geq0}}\frac{l b+nb^{-1}+\tfrac{Q}{2}-ix}{l b+nb^{-1}+\tfrac{Q}{2}+ix}\, ,
\ee
where $Q=b+b^{-1}$ and $b$ is the squashing parameter.  The partition function of an $\mathcal{N}=2$ theory with gauge group $G$ and $N_f$ chiral multiplets  is given by the following  integral over the Coulomb branch parameter $\sigma$: 
\ben
\label{genpart}
Z=\frac{1}{|W|}\int \prod_{j=1}^{r_G} d\sigma_j \, e^{2\pi i \xi \Tr(\sigma)} \, e^{\pi i k \Tr(\sigma^2)}
\frac{\prod_{a=1}^{N_f}  
 s_b\big( \tfrac{iQ}{2} (1-r_a)-\rho_a(\sigma)-\phi_a(M) \big)   }
{\prod_{\alpha } s_b \big(\tfrac{iQ}{2}\pm  \alpha(\sigma) \big)    } \;,
\een
where we used the shorthand notation $s_b(a\pm b) = s_b(a+b) \, s_b(a-b)$. In (\ref{genpart}) $|W|$ is the order of the Weyl group, $\alpha$ are the roots of the gauge group,  $\rho_a, \phi_a$ are the weights of the representations of the gauge and flavor groups.
We also introduced the $R$-charges $r_a$  and the real masses $M$  for the flavor symmetry.
The quadratic exponential is the contribution to the partition function of a level $k$ Chern-Simons coupling.
In the presence of $U(1)$ factors one can also turn on the  Fayet-Iliopoulos coupling  $\xi$.

\subsection[$U(N_c)$ with $N_f$ flavors and $\CW=\M^++\M^-$, and its dual]{$\boldsymbol{U(N_c)}$ with $\boldsymbol{N_f}$ flavors and $\boldsymbol{\CW=\M^++\M^-}$, and its dual}
\label{s3deriv}

In this section we check  at the level of the $S^3_b$ partition function the derivation of the duality $\CT_\M=\CT'_\M$ from the 4D Intriligator-Pouliot duality discussed in Section \ref{4d3dflow}. The  compactified Intriligator-Pouliot duality discussed  in \cite{Aharony:2013dha} relates  $\CT_1$ the  $USp(2N_c)$ theory with $2N_f$ fundamental  flavors  and monopole superpotential $\CW_1=\M$ and   $\CT_2$  the $USp(2N_f-2N_c-4)$ theory with $2N_f$ fundamental  flavors  and superpotential $\CW_2=\sum_{a<b} M^{ab} q_a q_b+ \widehat\M$. 
At the level of partition functions the duality is expressed by the equality  $Z_1=Z_2$ where:
\be
Z_1=\frac{1}{2^{N_c} N_c!}\int \prod_{j=1}^{N_c} d\sigma_j \,
\frac{\prod_{j=1}^{N_c} \prod_{a=1}^{N_f}  
 s_b\big(\tfrac{iQ}{2}\pm\sigma_j- m_a \big)   \prod_{b=1}^{N_f}   s_b \big(\tfrac{iQ}{2}\pm\sigma_j-\tilde m_b\big) }
{
\prod_{i<j}^{N_c} s_b \big(\tfrac{iQ}{2}\pm  (\sigma_j+\sigma_i) \big) \, s_b\big(\tfrac{iQ}{2}\pm  (\sigma_j-\sigma_i)\big)
\prod_{j=1}^{N_c}s_b\big(\tfrac{iQ}{2}\pm  2\sigma_j\big)}\;,
\ee
where we turned  on real masses $(m_1,\cdots m_{N_f},\tilde m_1,\cdots \tilde m_{N_f})$ for the flavor symmetry.
The monopole superpotential imposes the constraint:
\be
\label{cond} \sum_{a=1}^{N_f}m_a+\sum_{b=1}^{N_f}\tilde m_b=iQ (N_f-N_c-1) \;.
\ee
The partition function of the dual theory is 
\begin{multline}
Z_2 = \frac{1}{2^{N_c'} N_c'!}
\prod_{a<b}^{N_f} s_b \big(\tfrac{iQ}{2}-(m_a+m_b) \big) \, s_b\big(\tfrac{iQ}{2}-(\tilde m_a+\tilde m_b )\big)
\prod_{a,b=1}^{N_f}s_b\big(\tfrac{iQ}{2}-(m_a+\tilde m_b) \big) \\
\times \int \prod_{j=1}^{N_c'} d\sigma_j \,
\frac{\prod_{j=1}^{N_c'} \prod_{a=1}^{N_f}    s_b(m_a\pm\sigma_j)
 \prod_{b=1}^{N_f} s_b(\tilde m_b\pm\sigma_j)  }
{
\prod_{i<j}^{N_c'} s_b \big(\tfrac{iQ}{2}\pm  (\sigma_j+\sigma_i) \big) \, s_b\big(\tfrac{iQ}{2}\pm  (\sigma_j-\sigma_i)\big)
\prod_{j=1}^{N_c'}s_b\big(\tfrac{iQ}{2}\pm 2\sigma_j\big)  } \;,
\end{multline}
where $N_c' = N_f - N_c - 2$.
Now we consider a limit on the real mass deformation
\be
m_i \to m_i +s \,, \qquad \tilde m_i \to  \tilde m_i -s \,, \qquad i=1\,,\cdots, N_f\,,
\ee
with  $s\to \infty$ and focus on the vacuum corresponding to the saddle point at infinity. To do so we first observe that since the  integrands are symmetric  we can rewrite the integrals as: 
\be
\int_{-\infty}^{+\infty}\prod_{i=1}^{N_c} d\sigma_i  \, f(\sigma_i)= 2^{N_c} \int_{0}^{+\infty}\prod_{i=1}^{N_c} d\sigma_i \, f(\sigma_i)=
2^{N_c} \int_{-s}^{+\infty}\prod_{i=1}^{N_c} dx_i \, f(x_i+s) \;.
\ee
The matter contribution to the electric integrand is given by
\begin{multline}
\label{mat}
\prod_{j=1}^{N_c} \prod_{a=1}^{N_f}  s_b \big(\tfrac{iQ}{2}+x_i-m_a \big)  \prod_{b=1}^{N_f}   s_b\big(\tfrac{iQ}{2}-x_i-\tilde m_b\big) \\
\times
\prod_{j=1}^{N_c} \prod_{a=1}^{N_f}   s_b\big(\tfrac{iQ}{2}-x_i-m_a-2s \big) \prod_{b=1}^{N_f}   s_b\big(\tfrac{iQ}{2}+x_i-\tilde m_b+2s\big)\;,
\end{multline}
the limit splits the $2N_f$ chirals in the fundamental of $USp(2N_c)$ into a finite part corresponding to $N_f$ chirals in the fundamental of $U(N_c)$ and $N_f$ chirals in the anti-fundamental of $U(N_c)$. The remaining flavors have infinite mass and can be integrated out.
Similarly the vector multiplet contribution splits into a massless part coinciding with the  $U(N_c)$ vector multiplet contribution and two extra massive parts:
\be
\label{vec}
\prod_{i<j}^{N_c} s_b\big(\tfrac{iQ}{2}\pm  (x_j-x_i)\big)
\prod_{i<j}^{N_c} s_b\big(\tfrac{iQ}{2}\pm  (x_j+x_i)\pm 2s\big) \prod_{j=1}^{N_c}s_b\big(\tfrac{iQ}{2}\pm 2x_j\pm 2s\big) \;.  
\ee
To take the limit we use the  asymptotic behavior:
 \be
\label{asym}
\lim_{x\to \pm  \infty}s_b(x) \,\sim\, e^{\pm i\pi \frac{x^2}{2}} \;,
\ee
and find 
\begin{multline}
\label{alim}
\lim_{s\to\infty}Z_1= \frac{1}{ N_c!} \, e^{-\pi  s Q   N_c(  N_c+1)} \,
e^{\tfrac{i\pi}{2} \left(  N_c\sum_{a=1}^{N_f} \left( \tm_a^2-m_a^2+ i Q (m_a-\tm_a) \right) \right)} \\
\times\int \prod_{j=1}^{N_c} dx_j~
\frac{\prod_{j=1}^{N_c} \prod_{a=1}^{N_f}   s_b\big(\tfrac{iQ}{2}+x_j- m_a\big) \prod_{b=1}^{N_f}  s_b\big(\tfrac{iQ}{2}-x_j-\tilde m_b\big) }
{
\prod_{i<j}^{N_c}s_b\big(\tfrac{iQ}{2}\pm  (x_j-x_i)\big) }\;.
\end{multline}
To simplify  the exponential prefactor we used the following relation: \\
$\sum^{N_c}_{i<j}(2s+x_i+x_j) =  N_c(N_c-1)  s +  (N_c-1) \sum_{j=1}^{Nc} x_j$ and imposed the condition  (\ref{cond}).

\

The limit on the magnetic side produces a finite contribution to the integrand which can be identified with that of a $U(N_c'=N_f-N_c-2)$ theory with $N_f$ fundamental flavors:
\begin{multline}
\label{blim}
\lim_{s\to\infty}Z_2 = \frac{F_M}{ N_c'!} \, e^{-\pi  s Q   N'_c(  N'_c+1)} \,
e^{\tfrac{i\pi}{2}   N'_c\sum_{a=1}^{N_f}\left( m_a^2-\tm_a^2 \right)} \\
\times \int \prod_{j=1}^{N_c'} dx_j \,
\frac{\prod_{j=1}^{N_c'} \prod_{a=1}^{N_f}   s_b(m_a+x_j)
 \prod_{b=1}^{N_f}  s_b(\tilde m_b-x_j) }
{
\prod_{i<j}^{N_c'}s_b\big(\tfrac{iQ}{2}\pm  (x_j-x_i)\big)
}\;.
\end{multline}
We also have the the contribution of the  singlets:
\begin{multline}
F_M = \lim_{s\to\infty}\prod_{a<b}^{N_f} s_b\big(\tfrac{iQ}{2}-(m_a+m_b)-2s\big) \, s_b\big(\tfrac{iQ}{2}-(\tilde m_a+\tilde m_b )+2s\big)
\prod_{a,b=1}^{N_f}s_b\big(\tfrac{iQ}{2}-(m_a+\tilde m_b) \big) \\
= \prod_{a,b=1}^{N_f} s_b\big(\tfrac{iQ}{2}-(m_a+\tilde m_b) \big) \,
e^{-\tfrac{i\pi}{2} \sum_{a<b}(m_a+m_b+\tm_a+\tm_b-i Q) (m_a+m_b-\tm_a-\tm_b) } \\
\times e^{\pi Q s (N_f-1)(N_f-2N_c-2)}\;,
\end{multline}
which can be further simplified using that:
\begin{multline}
\sum_{a<b}(m_a+m_b+\tm_a+\tm_b-i Q) (m_a+m_b-\tm_a-\tm_b) = \\
= -iN_c\sum^{N_f}_{a=1} (m_a-\tm_a) + (N_f-2)\sum^{N_f}_{a=1} (m^2_a-\tm_a^2) \;.
\end{multline}
When we equate $(\ref{alim})=(\ref{blim})$ the divergent exponential prefactors, the dominant contributions to the saddles on the two sides, are equal and cancel out:
\be
\lim_{s\to\infty}Z_1\sim e^{-\pi  s Q   N_c(  N_c+1)}   =  e^{\pi Q s (N_f-1)(N_f-2N_c-2)}  e^{-\pi  s Q   N'_c(  N'_c+1)}   \sim
\lim_{s\to\infty} Z_2 \;.
\ee
We are then sure that we are comparing the same vacuum on the two side of the duality. The finite prefactors cancel-out too and in the end the equality $(\ref{alim})=(\ref{blim})$ yields:
\bea
\label{nonabel}
Z_{\CT_\M} &= \frac{1}{ N_c!}   
\int \prod_{j=1}^{N_c} dx_j \,
\frac{\prod_{j=1}^{N_c} \prod_{a=1}^{N_f}   s_b\big(\tfrac{iQ}{2}+x_j- m_a\big) \prod_{b=1}^{N_f}  s_b\big(\tfrac{iQ}{2}-x_j-\tilde m_b\big) }
{
\prod_{i<j}^{N_c}s_b\big(\tfrac{iQ}{2}\pm  (x_j-x_i)\big)
} \\
&= \frac{1}{ N_c'!}   
\prod_{a,b=1}^{N_f}s_b\big(\tfrac{iQ}{2}-(m_a+\tilde m_b) \big) \\
&\quad\times
\int \prod_{j=1}^{N_c'} dx_j \,
\frac{\prod_{j=1}^{N_c'} \prod_{a=1}^{N_f}   s_b(m_a+x_j)
 \prod_{b=1}^{N_f}  s_b(\tilde m_b-x_j) }
{
\prod_{i<j}^{N_c'}s_b\big(\tfrac{iQ}{2}\pm  (x_j-x_i)\big)
} = Z_{\CT'_\M} \;.
\eea
We identified  $Z_{\CT_\M}$ as the partition function of $\CT_\M$, the $U(N_c)$ theory with $N_f$  flavors and  $\CW= \M^++\M^-$ potential which breaks $U(1)_{T} \times U(1)_{A}$. Indeed there are no real masses turned on for these symmetries since there is no FI term and the masses satisfy the constraint  (\ref{cond}).  Similarly we identify $Z_{\CT'_\M}$ as the partition function of the  dual theory  $\CT'_\M$, the $U(N_f-N_c-2)$ theory with $N_f$ flavors and  $\CW=  \widehat\M^++\widehat\M^-+ \sum_{a,b}^{N_f} M\ud{a}{b}  \tilde q_a q^b$. Indeed we also see the contribution of the $N_f^2$  singlets $M\ud{a}{b}$ with masses $(m_a+\tilde m_b) $. This is test at the level of the partition function of our $\CT_\M=\CT'_\M$ duality.\\

For $N_f=N_c+2$ the integral in the magnetic theory disappears and, as expected from the discussion in Section \ref{WZM}, we find  $N_f^2$ chiral singlets interacting with superpotential $\det(M)$.\footnote{As observed in \cite{Willett:2011gp} the convergence of 3D partition functions is controlled by the dimensions of the fundamental monopoles. In particular the asymptotic behavior in the electric theory  is given by $e^{-R[\M^\pm]x}$ while in the magnetic theory it is given by $e^{-R[\widehat \M^\pm]x}$.
In our case, since the monopoles enter the superpotential and are exactly marginal, both $Z_{\CT_\M}$ and $Z_{\CT'_\M}$ remain convergent even when we enter the region $N_c + 3 \leq N_f \leq \frac{4}{3}(N_c+1)$ where the dual mesons become free. This has to be contrasted with the behavior discussed in  \cite{Safdi:2012re}.}

\subsection{Real mass deformation to  the Aharony duality}
\label{reaha}

A further consistency of the $\CT_\M=\CT'_{\M}$  duality is to show that it  reduces to the  Aharony duality with a suitable real mass deformation. We start with $N_f+2$ flavors and consider  the  following deformation: 
\ben
\label{ahadef}
\nn&&m_{N_f+1}\to  m_{N_f+1}  +t\;, \qquad  \tilde m_{N_f+1}\to  m_{N_f+1} -t \;, \\
&&m_{N_f+2}\to  m_{N_f+2}  -t\;, \qquad  \tilde m_{N_f+2}\to  m_{N_f+2} +t\;,
\een
with $t\to\infty$. By defining  $\eta=2 m_{N_f+1}+2m_{N_f+2}$ and $\xi=2m_{N_f+1}-2m_{N_f+2}$, the mass constraint (\ref{cond}) becomes:
\be
\label{ahacons}
\sum_{a,b=1}^{N_f}(m_a+\tilde m_b)+\eta=2\omega (N_f-N_c+1)\;.
\ee
Since $\eta$ is a free parameter, this constraint is lifted. By using the asymptotics (\ref{asym}) the limit of the electric side of the identity (\ref{nonabel})  with $N_f+2$ flavors, in the trivial vacuum, becomes:
\be
\label{ahaele}
\frac{e^{\pi i N_c t (2i Q- \eta)} }{N_c!} \int \prod_{j=1}^{N_c} dx_j \,
\frac{\prod_{j=1}^{N_c} \prod_{a=1}^{N_f}  e^{\pi i (\sum_j x_j)\xi} s_b \big(\tfrac{iQ}{2}+x_j- m_a\big) \prod_{b=1}^{N_f}  s_b\big(\tfrac{iQ}{2}-x_j-\tilde m_b\big) }
{
\prod_{i<j}^{N_c}s_b\big(\tfrac{iQ}{2}\pm  (x_j-x_i)\big)}\;.
\ee
Up to the divergent factor this is the partition function of a $U(N_c)$ theory with $N_f$ flavors  and $\CW=0$.
Indeed the constraint on the masses is lifted  and $\xi$ enters as an FI parameter, we have then restored the $U(1)_{T}\times U(1)_{A}$ symmetries.
By taking the same limit on the magnetic  side of   (\ref{nonabel}) we find: 
\begin{multline}
\label{ahamag}
e^{\pi i (N_f-N_c) t \eta} \,
e^{-2\pi i t \left(  \sum_{j=1}^{N_f}(m_j+\tilde m_j)  +(2+N_f)  ( m_{N_f+1} +m_{N_f+2})  -i Q (N_f+1) \right)}
\\
\times e^{\pi i    ( m_{N_f+1} -m_{N_f+2})\sum_{j}^{N_f}(m_j-\tilde m_j)}
 s_b\big(\tfrac{iQ}{2}-2m_{N_f+2} \big) s_b\big(\tfrac{iQ}{2}-2m_{N_f+1}\big)
\prod_{a,b=1}^{N_f}s_b\big(\tfrac{iQ}{2}-(m_a+\tilde m_b) \big) \\
\times
\frac{1}{(N_f-N_c)!} \int \prod_{j=1}^{N_f-N_c} dx_j \, e^{\pi i \xi \sum_j x_j} \,
\frac{\prod_{j=1}^{N_f-N_c} \prod_{a=1}^{N_f}   s_b(m_a+x_j)
 \prod_{b=1}^{N_f}  s_b(\tilde m_b-x_j) }
{
\prod_{i<j}^{N_f-N_c}s_b\big(\tfrac{iQ}{2}\pm  (x_j-x_i)\big)
} \;.
\end{multline}
The leading contributions to the saddle points in (\ref{ahaele}) and (\ref{ahamag})  when using (\ref{ahacons}) are equal and cancel out. The finite exponentials also simplify and cancel out. It is convenient to introduce the following parametrisation:
\be
\label{ahapar}
m_a=\mu_a-M_a\,,\qquad \tilde m_a=\mu_a+M_a\,, \qquad \sum\nolimits_a M_a=0\,,
\ee
with 
\be
\eta=i Q(N_f-N_c+1)-2 \sum_{a=1}^{N_f}\mu_a\,,
\ee
in the Cartan of the global flavor symmetry $SU(N_f)_{M_a} \times SU(N_f)_{m_a}\times U(1)_\eta$, which allows us to  rewrite  the equality of eqs. (\ref{ahaele}) and (\ref{ahamag})  as:
\begin{multline}
\label{aharec}
\frac{1}{N_c!} \int \prod_{j=1}^{N_c} dx_j \,
\frac{\prod_{j=1}^{N_c} \prod_{a=1}^{N_f}  e^{\pi i \xi \sum_j x_j} \, s_b\big(\tfrac{iQ}{2}\pm (x_j+M_a)- \mu_a\big) }
{
\prod_{i<j}^{N_c}s_b\big(\tfrac{iQ}{2}\pm  (x_j-x_i)\big)} = \\
= s_b\big(\tfrac{iQ   }{2}-\tfrac{iQ (N_f-N_c+1)   -2\sum_a\mu_a\pm \xi     }{2} \big)
\prod_{a,b=1}^{N_f}s_b\big(\tfrac{iQ}{2}-(\mu_a+\mu_b-M_a+M_b) \big) \\
\times
\frac{1}{(N_f-N_c)!}\int \prod_{j=1}^{N_f-N_c} dx_j \, e^{\pi i \xi \sum_j x_j} \,
\frac{\prod_{j=1}^{N_f-N_c} \prod_{a=1}^{N_f}   s_b\big(\pm(x_j-M_a)+\mu_a\big)
}
{
\prod_{i<j}^{N_f-N_c}s_b\big(\tfrac{iQ}{2}\pm  (x_j-x_i)\big)
}\;.
\end{multline}
By looking at the real masses appearing in the double sine functions on  the magnetic side, we see that besides the $N_f\times N_f$
singlets with masses $(\mu_a+\mu_b-M_a+M_b)$ there are  two extra singlets  with R-charge $(N_f-N_c+1)$ and topological charge $\pm 1$ which   can be  identified with the singlets $S^\pm$  entering the   superpotential of the dual theory  $\CW=\sum_{a,b}^{N_f}   M\ud{a}{b} \tilde q_a q^b +\widehat \M^+ S^-+ \widehat\M^- S^+$. We have thus shown that our duality reduces to the Aharony duality (\ref{aharec}).

\section{$\boldsymbol{U(N_c)}$ with $\boldsymbol{N_f}$ flavors and $\boldsymbol{\CW=\M^-}$ and its dual}
\label{newonemon}

In this section we  show that a suitable real mass deformation of the $\CT_\M=\CT'_{\M}$ duality allows us to derive a new duality involving the $U(N_c)$ theory with $N_f$ flavors with a  superpotential involving only one monopole operator $\CW= \M^\pm$. We start from $\CT_\M=\CT'_{\M}$  with $N_f+1$ flavors and consider the real mass deformation
\ben
\label{replu}
m_{N_f+1}\to  m_{N_f+1}  +t\,, \qquad  \tilde m_{N_f+1}\to  m_{N_f+1} -t\,,
\een
with $t\to\infty$. By  defining   $\eta=2 m_{N_f+1}$ the mass constraint (\ref{cond}) becomes
\be
\label{onemonacons}
\sum_{a,b=1}^{N_f}(m_a+\tilde m_b)+\eta=Q (N_f-N_c)\;,
\ee
and it is lifted since $\eta$ is a free parameter. By using the asymptotics (\ref{asym}), the limit of the electric side of the identity (\ref{nonabel}) in the trivial vacuum yields
\be
\label{elgonem}
\frac{e^{\pi i N_c t (i Q- \eta)} }{N_c!} \int \prod_{j=1}^{N_c} dx_j \,
e^{\pi i (\eta-i Q) \sum_j x_j} \frac{\prod_{j=1}^{N_c}  \prod_{a=1}^{N_f}  s_b\big(\tfrac{iQ}{2}+x_j- m_a\big) \prod_{b=1}^{N_f}  s_b\big(\tfrac{iQ}{2}-x_j-\tilde m_b\big) }
{
\prod_{i<j}^{N_c}s_b\big(\tfrac{iQ}{2}\pm  (x_j-x_i)\big)} \;.
\ee
Apart from the divergent exponential prefactor, the leading contribution to the saddle point is the partition function of a $U(N_c)$ theory with $N_f$ flavors and $\CW= \M^-$. This monopole superpotential removes  $\M^-$ but leaves $\M^+$ in the chiral ring and
breaks the $U(1)_{T}\times U(1)_{A}$ symmetry  to the diagonal,  indeed the FI parameter and the axial mass are not independent. By taking the same limit on the magnetic  side of  (\ref{nonabel}) we find: 
\begin{multline}
 \label{magonem}
e^{\pi i (N_f-N_c-1) t \eta} \, e^{\pi i t \left(-\sum_j^{N_f}(m_j+\tilde m_j)-\eta-(N_f-1)\eta+ i Q N_f \right)} 
\\
\times e^{\pi i \sum_j^{N_f} \left( \tfrac{(m_j^2-\tilde m_j^2)}{2} + (m_j-\tilde m_j)(\eta-i Q)  \right)} 
 s_b\big(\tfrac{iQ}{2}-\eta \big)
\prod_{a,b=1}^{N_f}s_b \big(\tfrac{iQ}{2}-(m_a+\tilde m_b) \big) \\
\times
\tfrac{1}{(N_f-N_c-1)!}\int \prod_{j=1}^{N_f-N_c-1} \!\!\! dx_j \, e^{\pi i \eta \sum_j x_j}
\frac{\prod_{j=1}^{N_f-N_c-1} \prod_{a=1}^{N_f}   s_b(m_a+x_j)
 \prod_{b=1}^{N_f}  s_b(\tilde m_b-x_j) }
{
\prod_{i<j}^{N_f-N_c-1}s_b\big(\tfrac{iQ}{2}\pm  (x_j-x_i)\big)
}\;.
\end{multline}
Importantly, the divergent prefactor in (\ref{magonem})---\ie{} the leading contribution to the magnetic saddle---equals the electric one in  (\ref{elgonem}):
\be
e^{\pi i N_c t (i Q- \eta)} =
e^{\pi i (N_f-N_c-1) t \eta} \, e^{\pi i t \left( -\sum_j^{N_f}(m_j+\tilde m_j)-\eta-(N_f-1)\eta+ i Q N_f \right)} \;,
\ee
and we are then sure that we are comparing the same vacuum on the two side of the duality.
Some finite exponential prefactors cancel out too. Again we can introduce  the  parameterisation (\ref{ahapar}) with the constraint becoming:
\be
\label{onemonacons2}
\eta=i Q(N_f-N_c)-2 \sum_{a=1}^{N_f}\mu_a\,,
\ee
and by equating  eqs. (\ref{elgonem}) and (\ref{magonem}) we arrive to the following identity:
\begin{multline}
\label{onemon}
Z_1= \frac{1}{N_c!} \int \prod_{j=1}^{N_c} dx_j \,
\frac{\prod_{j=1}^{N_c} \prod_{a=1}^{N_f}    e^{\pi i (\sum_j x_j)  (\eta-i Q)}  s_b\big(\tfrac{iQ}{2}\pm (x_j+M_a)- \mu_a\big) }
{
\prod_{i<j}^{N_c}s_b\big(\tfrac{iQ}{2}\pm  (x_j-x_i)\big)} \\
= e^{-2\pi i \sum_a^{N_f} M_a \mu_a } 
s_b\big(\tfrac{iQ  }{2}-\eta \big)
\prod_{a,b=1}^{N_f}s_b\big(\tfrac{iQ}{2}-(\mu_a+\mu_b-M_a+M_b) \big) \\
\times
\tfrac{1}{(N_f-N_c-1)!}\int \prod_{j=1}^{N_f-N_c-1} \!\!\! dx_j \, e^{\pi i \eta \sum_j x_j}
\frac{\prod_{j=1}^{N_f-N_c-1} \prod_{a=1}^{N_f}   s_b\big(\pm(x_j-M_a)+\mu_a\big)
}
{
\prod_{i<j}^{N_f-N_c-1}s_b\big(\tfrac{iQ}{2}\pm  (x_j-x_i)\big)
} = Z_2 \;.
\end{multline}
On the dual side we have a  $U(N_f-N_c-1)$ theory  with $N_f$ flavors. We also have $N_f^2$ singlets $M_{ab} $ with masses  $(\mu_a+\mu_b-M_a+M_b)$ and a singlet  $S^+$ whose contribution to the partition function is:
 \be
s_b\big(\tfrac{iQ  }{2}-\eta \big)=s_b\big(\tfrac{iQ  }{2}-\tfrac{\eta}{2} -\tfrac{\eta}{2} \big)=s_b\big(\tfrac{iQ  }{2}-\tfrac{i Q(N_f-N_c)-2 \sum_a^{N_f}\mu_a}{2} -\tfrac{\eta}{2} \big) \;.
\ee
From here we see that $S^+$ has topological  charge $1$ and R-charge $N_f-N_c$  and enters the superpotential as   $S^+ \widehat\M^-$.
Also on the dual side we see that  the $U(1)_{T}\times U(1)_{A}$ symmetry is broken to the diagonal,  indeed we still have the constraint (\ref{onemonacons2})  relating the  FI parameter $\eta$ and the total axial mass $\sum_a \mu_a$. This is consistent with the presence of a linear monopole term in the superpotential. 

So we propose that the  duality expressed by  the identity (\ref{onemon}) relates 
\be
\CT_1 : \qquad U(N_c) \text{~~SQCD with $N_f$ flavors} \;, \CW =\M^-
\label{t1mp}\ee
and
\ben
\nonumber \CT_2 : \qquad & U(N_f - N_c - 1) \text{~~SQCD with $N_f$ flavors, $N_f^2$ singlets} \;,&\\
& \CW = \widehat\M^+ + \widehat\M^- S^+ + \sum_{i,j=1}^{N_f} M\ud{i}{j} \tilde q_i q^j \;. &
\label{t2mp}
\een

An analogous duality  can be obtained by taking the real mass deformation (\ref{replu}) with $t\to-\infty$: it relates $U(N_c)$ with $N_f$ flavors and $\CW =\M^+$ to $U(N_f - N_c - 1)$ with  $N_f$ flavors  and $ \CW = \widehat\M^- + \widehat\M^+ S^- + \sum_{i,j=1}^{N_f} M\ud{i}{j} \tilde q_i q^j$.

\

In the case $N_f=N_c+1$ the dual theory has no integration on the dynamical variables so the dual theory has no gauge group and we propose that it can be described by  $N_f^2+1$ singlets interacting with $W= S^+ \det(M)$.
This duality for  $N_c=1, N_f=2$ has been discussed in \cite{Collinucci:2016hpz}.

\subsection{Chiral real mass deformation: Chern-Simons theories}
\label{chiralsec} 

We can generate dualities with Chern-Simons couplings starting from the duality with one monopole (\ref{t1mp})-(\ref{t2mp}) and considering suitable  real mass deformations.\footnote{The Chern-Simons coupling induces a gauge charge for one of the two fundamental monopoles and only one gauge-invariant fundamental monopole can be added to the superpotential.}
For example in (\ref{onemon}) we can consider the deformation
\begin{eqnarray}
\nonumber &\mu_a\to\mu_a\,, \quad M_a\to M_a+s\,, \quad a \neq i\,,  \quad  M_i\to M_i-(N_f-1) s\,, \quad \mu_i\to\mu_i+ N_f s\,,&\\&
\eta \to \eta-2N_fs\,,\qquad s\to\infty\,,& 
\end{eqnarray}
which preserves the constraint (\ref{onemonacons2}).
On the electric side we consider the vacuum where one of the chirals gets a large mass, to do so we shift the Cartan variables $x_j\to x_j -s$ so that only the $i$-th chiral can be integrated out and generates a half Chern-Simons coupling:
\be
s_b\big(\tfrac{iQ}{2}+x_j+M_i-\mu_i-2 N_f s\big) \,\to\,  e^{-\tfrac{i\pi}{2} \left( \tfrac{iQ}{2}+ x_j+M_i-\mu_i-2 N_f s \right)^2} \;.
\ee
Taking the $s\to \infty$ limit on the electric side of (\ref{onemon}) yields a divergent prefactor $e^{A s^2 + B s }$, where $A$ is just a numerical coefficient depending on $N_f,N_c$ and $B$ a linear combination of the mass parameters and a finite part:
\begin{multline}
\label{elecs}
\lim_{t\to\infty}Z_1 \,\sim\, \frac{e^{A s^2 + B s }}{N_c!}   
\int \prod_{j=1}^{N_c} dx_j \, e^{-\tfrac{i\pi}{2} (\sum_j x_j^2)} \,
e^{\pi i (\sum_j x_j)  \left( \eta-M_i+\mu_i-\tfrac{3i Q}{2} \right)} \\
\times \frac{\prod_{j=1}^{N_c}  
s_b\big(\tfrac{iQ}{2}- x_j-M_i-\mu_i\big)     \prod_{a\neq i }^{N_f}   s_b\big(\tfrac{iQ}{2}\pm (x_j+M_a)- \mu_a\big) }
{
\prod_{i<j}^{N_c}s_b\big(\tfrac{iQ}{2}\pm  (x_j-x_i)\big)} \;.
\end{multline}
The coefficient $e^{A s^2 + B s }$ is the leading contribution to the saddle point of the electric partition function we are focusing on and we need to single out the same vacuum with the same leading contribution on the magnetic side. This can be done by shifting $x_j \to x_j+s$ on the magnetic side so that the $i$-th chiral is integrated out generating a half Chern-Simon coupling with opposite sign:
\be
s_b\big(x_j-M_i+\mu_i+2N_f s\big) \,\to\, e^{\tfrac{i\pi}{2} (x_j-M_i+\mu_i+2 N_f s)^2}\;.
\ee
The leading contribution to saddle point now gets contributions also from the exponential prefactors and from the singlets on the magnetic side of (\ref{onemon}). By a tedious but straightforward computation we obtain the same divergent coefficient $e^{A s^2 + B s }$ we found on the electric side multiplying a finite part:
\begin{multline}
\label{magcs}
\lim_{t\to\infty}Z_2 \,\sim\, e^{A s^2 + B s }\,\prod_{a\neq i}s_b\big(\tfrac{iQ}{2}-(\mu_a+\mu_i-M_a+M_i) \big)\prod_{a,b\neq  i}^{N_f}  s_b\big(\tfrac{iQ}{2}-(\mu_a+\mu_b-M_a+M_b) \big) \\
\frac{1}{(N_f-N_c-1)!}
\int \prod_{j=1}^{N_f-N_c-1} dx_j  \, e^{\tfrac{i\pi}{2} (\sum_j x_j^2)} \, e^{\pi i (\sum_j x_j)(\eta-M_i+\mu_i)} \\
\frac{\prod_{j=1}^{N_f-N_c-1}  s_b(-x_j+M_i+\mu_i) \prod_{a\neq i}^{N_f}   s_b\big(\pm(x_j-M_a)+\mu_a\big)}
{
\prod_{i<j}^{N_f-N_c-1}s_b\big(\tfrac{iQ}{2}\pm  (x_j-x_i)\big)
} \;.
\end{multline}
In conclusion, since the electric and magnetic saddles that we have singled out have the same leading asymptotics, we propose to compare those two isolated vacua. This gives a new duality%
\footnote{We are omitting some finite exponential coefficients, the contribution of background Chern-Simons terms.}
expressed by  the identity $(\ref{elecs})=(\ref{magcs})$ which relates:
\be
U(N_c)_{\frac12} \text{ with } (N_f, N_f-1) \text{ fund/antifund chirals} \;,\; \CW =\M^-
\ee
and
\ben
\nonumber  & U(N_f - N_c - 1)_{-\frac12} \text{ with } (N_f, N_f-1) \text{ fund/antifund, $N_f(N_f-1)$ singlet chirals},&\\
& \CW = \widehat\M^+  + \sum_{i}^{N_f}\sum_{j}^{N_f-1} M\ud{i}{j} \tilde q_i q^j \;.
\een
The linear terms in the monopole break $U(1)_A\times U(1)_T$ to the diagonal. Indeed the constraint (\ref{onemonacons2})
relating the FI and the axial mass is still preserved.

This duality can be easily generalized by giving mass to $2k$  charge $-1$ chirals, leading to the family of dualities:
\be
U(N_c)_{\frac k2} \text{ with } (N_f, N_f-k) \text{ fund/antifund chirals} \;,\; \CW =\M^-
\ee
and
\ben
\nonumber  & U(N_f - N_c - 1)_{- \frac k2} \text{ with } (N_f, N_f-k) \text{ fund/antifund, $N_f(N_f-k)$ singlet chirals},&\\
& \CW = \widehat\M^+  + \sum_{i}^{N_f}\sum_{j}^{N_f-k} M\ud{i}{j} \tilde q_i q^j
\een
which provide a generalization with monopole superpotential of the dualities discussed in \cite{Benini:2011mf}.

\section*{Acknowledgements}
We are grateful to Matthew Buican and Diego Rodriguez-Gomez for comments on the draft. We also thank Claudio Destri for help with numerical computations. F.B. was supported in part by the MIUR-SIR grant RBSI1471GJ ``Quantum Field Theories at Strong Coupling: Exact Computations and Applications'', by the INFN, and by the IBM Einstein Fellowship at the Institute for Advanced Study.
S.B. is partly supported by the INFN Research Projects GAST and ST$\&$FI and by PRIN ``Geometria delle variet\`a algebriche".
S.P. is partially  supported by the  ERC-STG grant 637844-HBQFTNCER.

 \appendix

\section{The Abelian case}\label{abeliansection}

In the Abelian case, $N_c=1$, we have at our disposal Abelian mirror symmetry \cite{Aharony:1997bx, Kapustin:1999ha}, and in the mirror it is easier to understand the effect of the monopole deformation. 
The mirror of the SQED with $N_f$ flavors and $\CW=0$ \cite{Aharony:1997bx} is $\CT^\text{mirror}_0$, a $U(1)^{N_f-1}$ gauge theory with $3N_f$ chiral multiplets $\Phi_i$, $q_i$, $\tilde q_i$, $i=1, \dots, N_f$, described by the quiver diagram
$$
\begin{tikzpicture}
\path (0,0) node[rectangle,draw](x1) {$1$} -- (2,0) node[circle,draw](x2) {$1$} -- (4,0) node[circle,draw=white](x3) {$\dots$} -- (6,0) node[circle,draw](x4) {$1$} -- (8,0) node[rectangle,draw](x5) {$1$};
\draw [->] (x1) to[bend left] (x2); \draw [<-] (x1) to[bend right] (x2);
\draw [->] (x2) to[bend left] (x3); \draw [<-] (x2) to[bend right] (x3);
\draw [->] (x3) to[bend left] (x4); \draw [<-] (x3) to[bend right] (x4);
\draw [->] (x4) to[bend left] (x5); \draw [<-] (x4) to[bend right] (x5);
\draw [->] (x1) to[out=120, in=180] (0,1) to[out=0,in=60] (x1);
\draw [->] (x2) to[out=120, in=180] (2,1) to[out=0,in=60] (x2);
\draw [->] (x4) to[out=120, in=180] (6,1) to[out=0,in=60] (x4);
\node at (x3) {$\underbrace{\rule{12em}{0pt}\rule{0pt}{4.5em}}_{N_f-1}$};
\end{tikzpicture}
$$
and with superpotential $\CW = \sum_{i=1}^{N_f} \Phi_i q_i \qt_i$. 

In order to reach the mirror of $\CT_\M$, that we call $\CT^\text{mirror}_\M$, we add to the superpotential the mirror dual of the chiral operator $\CW_\text{\rm mon}$, namely $\prod_i q_i + \prod_i \qt_i$. The full superpotential of $\CT^\text{mirror}_\M$ reads
\be
\label{abelianmirrorM}
\CW^\text{mirror}_\M = \sum_{i=1}^{N_f} \Phi_i q_i \qt_i + \prod_{i=1}^{N_f} q_i + \prod_{i=1}^{N_f} \qt_i \;.
\ee

For $N_f \leq 3$ the operator $\CW_\text{mon}$ is relevant, while for $N_f > 3$ the $\CW_\text{mon}$ is irrelevant. This is true, obviously, both in $\CT_0$ and in $\CT^\text{mirror}_0$. Yet, it is possible to go beyond $N_f = 3$ in the mirror. In fact it is possible to reach $\CT^\text{mirror}_\M$ by first turning on the $N_f-1$ gauge couplings (this step decreases the R-charges of the $2N_f$ charged chiral fields), then turning on $\CW_\text{mon} = \prod q_i + \prod \qt_i$, and as a third step coupling the mesons $q_i \qt_i$ to the singlets $\Phi_i$. An $\cF$-maximization computation shows that the second step is possible as long as $N_f < 6$.%
\footnote{One can see analytically that it is impossible for $\CW_\text{mon} = \prod q_i + \prod \qt_i$ to be relevant for $N_f \geq 6$. The $U(1)^{N_f-1}$ gauge theory with gauge couplings turned on and no singlets is mirror to a $U(1)$ gauge theory with $N_f$ flavors $Q_i, \Qt_j$, $N_f$ singlets $\phi_i$ and $\CW= \sum_{i=1}^{N_f} \phi_i Q_i \Qt_i$. This theory can be thought of a collection of $N_f$ XYZ models coupled by a $U(1)$ gauge field. The R-charges of the XYZ models are all $\frac{2}{3}$, then adding a gauge interaction decreases a bit the R-charges of the $2N_f$ charged fields $Q_i,\Qt_i$ while increasing the R-charge of the singlets $\phi_i$. Under the mirror map, the singlets $\phi_i$ are mapped to the mesons $q_i \qt_i$ in the $U(1)^{N_f-1}$ theory. Hence the mesons have dimension higher than $\frac23$, $\Delta[q_i \qt_i] > \frac23$, and so $\Delta \big[ \prod q_i + \prod \qt_i \big] > \frac{N_f}{3}$, which for $N_f \geq 6$ is certainly beyond the relevance bound $\Delta=2$.}
From $\cF$-maximization we obtain the dimensions
\be
\begin{array}{c|cccc}
N_f & 2 & 3 & 4 & 5 \\
\hline\hline
\rule[0em]{0pt}{1.2em} \Delta \big[ \prod q_i + \prod \qt_i \big] \; & \; 0.82 \; & \; 1.14 \; & \; 1.48 \; & \; 1.81 \;
\end{array}
\ee
So in this range we expect that $\CT^\text{mirror}_\M$ with superpotential (\ref{abelianmirrorM}) can be reached from the free theory without adding additional degrees of freedom.

The mesonic chiral ring (or Higgs branch) of the mirror, given by  gauge invariant operators built out of the $q_i$, $\qt_i$, consists at most of a single point:  the F-term relations set $q_i \tilde q_i = 0$ as well as $\prod_i q_i = \prod_i \tilde q_i = 0$. Thus the chiral ring of the mirror is generated by $\Phi_i$ and the BPS monopole operators only. In terms of the original theory $\CT_\M$ with $N_c=1$, it means that we only have the Higgs branch as expected.

For $N_f = N_c = 1$, the mirror is simply the Wess-Zumino model with superpotential $\CW^\text{mirror}_{N_c = N_f = 1} = \Phi q \tilde q + q + \tilde q$ with no supersymmetric vacua. If $N_f =2$, the mirror is a $U(1)$ theory with two flavors and two extra singlets, and superpotential given by $\CW^\text{mirror} = \Phi_1 q_1 \tilde q_1 + \Phi_2 q_2 \tilde q_2 + m (q_1 q_2 + \tilde q_1 \tilde q_2)$. In particular the monopole deformation appears in the mirror as a mass deformation, and we have introduced a coupling $m$. The Coulomb branch of this theory \cite{Aharony:1997bx} is described by the quantum relation
\be
\fN_+ \fN_- = \det \cM = \Phi_1 \Phi_2 - m^2 \;,
\ee
where $\fN_\pm$ are the monopoles of the mirror theory and $\cM = \smat{ \Phi_1 & m \\ m & \Phi_2}$ is the mass matrix of the quarks. Given the identifications $\Phi_i = M\ud{i}{i}$, $\fN^+ = M\ud{1}{2}$ and $\fN^- = M\ud{2}{1}$ (where $M\ud{i}{j}=q^i \tilde q_j$) through mirror symmetry, in terms of the variables of the original theory that is
\be
\det M = m^2 \;,
\ee
which is a quantum deformation of the classical Higgs branch $\det M = 0$. This is in agreement with  the results of Section \ref{sec: complex mass def}.

\bibliographystyle{JHEP}
\bibliography{Wmonopole}
\end{document}